**Working Paper**

# "Stabilizer" or "catalyst"? How green technology innovation affects the risk of stock price crashes: an analysis based on the quantity and quality of patents

*Ge-zhi Wu*[1,2] | *Da-ming You*[1,2]

[1] Business School, Central South University,Changsha, China

[2] Collaborative Innovation Center of Resource-conserving and Environmentally friendly Society and Ecological Civilization, Central South University, Changsha, China

**Correspondence**

Ge-zhi Wu, Business School, Central South University; Collaborative Innovation Center of Resource-conserving and Environmentally friendly Society and Ecological Civilization, Central South University, No. 932 Lushan South Road, Yuelu District, Changsha, Hunan 410083, China

Personal Email: 649792656@qq.com

Institutional Email：wgzz2218@csu.edu.cn

Telephone number: 86-18975833865

**Funding information**

National Natural Science Foundation of China, Grant/Award Numbers: 71974209;

**ACKNOWLEDGMENTS**

The authors acknowledge the anonymous reviewers and editors for helpful guidance on prior versions of the article.

## ABSTRACT

To explore the relationship between corporate green technological innovation and the risk of stock price crashes, we first analyzed the data of listed companies in China from 2008 to 2018 and constructed indicators for the quantity and quality of corporate green technology innovation. The study found that the quantity of green technology innovation is not related to the risk of stock price crashes, while the quality of green technology innovation is negatively related to the risk of stock price crashes. Second, we studied the impact of corporate ownership on the relationship between the quality of green technological innovation and the risk of stock price crashes and found that in nonstate-owned enterprises, the quality of green technological innovation is negatively correlated with the risk of a stock price collapse, while in state-owned enterprises, the quality of green technological innovation and the risk of a stock price collapse are positive and not significant. Furthermore, we studied the mediating effect of the number of negative news reports in the media of listed companies on the relationship between the quality of corporate green technology innovation and the stock price crash and found that the quality of green technology innovation is positively correlated with the number of negative news reports in the media of listed companies, while the number of negative news reports in the media of listed companies is positively correlated with the risk of a stock price collapse. Finally, we conducted a DID regression by using the impact of exogenous policy shocks on the quality of green technology innovation, and the main results passed the robustness test.

## 1| INTRODUCTION

Sustainable development is an important global issue. To achieve the 2030 sustainable development goals proposed by the United Nations, countries around the world must curb ecosystem degradation, reduce environmental pollutant emissions, and achieve coordinated development of the environment and economy. The top two economies in the world have both increased their emphasis on green development; in the United States, the Biden administration returned to the Paris Agreement, and the green economy may become a new engine of future economic development and of overcoming the economic crisis in the postepidemic era, while in China, the government put forward the green development goal of peaking carbon emissions by 2030 and carbon neutrality by 2060, which demonstrates the great importance of environmental issues to the Chinese government and people. What means and measures should be taken to ensure that green development runs along the optimal path is an urgent and important issue(Tong et al., 2016).

Green technology innovation is usually one of the important measures to respond to and realize green development. Green technology refers to the general technologies, processes and products that follow ecological principles and development laws, save resources and energy consumption, and avoid or reduce environmental pollution and damage. However, since the investment in and return on green technology innovation are inherently uncertain, are characterized by dual externalities of public goods of the environment and innovation and are affected by many factors, including government regulations as well as corporate, technical, and ecological factors, motivating relevant companies to take a certain degree of risk and actively carry out green innovation activities is a key challenge. Scholars have explored the driving factors of green technology innovation from many perspectives (Christmann et al., 2000; Claudiy et al., 2016; BU et al., 2016;Scandelius et al., 2018; Juntunen et al., 2019; Cainelli et al., 2020).

The influencing factors of stock price crashes are an important issue worthy of study. For investors, stock price crashes are a small probability event, yet asset management models are typically based on the law of large numbers, making them lose their predictive ability and potentially causing huge losses for related investors. Therefore, it is important to understand the causes of the risk of stock price collapse. The current research on stock price crashes is mainly conducted from the perspectives of management and CEOs, such as CEO age (Andreou et al., 2017) and CEO duality (Chen et al., 2015). Other studies mainly focus on management behavior and investment and external supervisors.

The capital market is an important place to reflect and judge the value of a company. Corporate green technological innovation activities should be reflected in the capital market. At present, some scholars have studied the relationship between corporate social responsibility and stock price collapse risk (Kim et al., 2014; Zhang, et al., 2016). However, the relationship between green technology innovation and the risk of a stock price collapse is not considered. Is corporate green technological innovation a "stabilizer" or a "catalyst" for the risk of a stock price collapse? Has the risk of a stock price collapse been reduced? Whether the enterprise's green technology innovation behavior gets positive feedback from the market will be a problem worthy of study. If the company's green technology innovation behavior can get positive feedback in the risk assessment of stock price collapse, it will help to stimulate the enterprise's green technology innovation.

Whether the impact of Chinese corporate green technology innovation on the risk of a stock price collapse is more "instrumental(tool)" or "value-adding" still needs further study. At the same time, a series of questions are posed. Under the characteristics of dual externalities of green technology innovation, in which the design of the current Chinese institutional system is far from perfect and the market regulatory policies still need to be improved, what is the impact of corporate green technology innovation on the risk of a stock price collapse? Going deep into the internal operating mechanism of related companies, does corporate green technology innovation embody a "self-interested tool" to conceal management's failures and improper behavior or a "value tool" to enhance shareholders' wealth? What is the internal mechanism of action? These questions motivate this research study.

Many studies on stock price crashes are based on relevant data from U.S. companies. However, we choose Chinese listed companies as research subjects. First, China is facing a more serious environmental pollution problem than the West; thus, green innovation has more urgent needs and value in China. Relevant data show that China’s comprehensive national strength ranks second in the world, but its environmental governance ability ranks second from the bottom, and in the comprehensive environmental quality index compiled by Yale University, China ranks 109th in the world in environmental quality. Second, compared with American companies, the Chinese corporate system has its own unique characteristics, including a relatively low proportion of institutional investors’ shareholding, government control over some companies, and less efficient independent directors. Third, overall, the volatility of the Chinese stock market is higher than that of the U.S. stock market; according to our calculations, in the sample period, the standard deviation of the monthly stock index yields of the Chinese stock market is approximately twice the standard deviation of the S&P 500 index yields. Studying the crash risk of the Chinese stock market is complex and difficult because the stock market is heavily influenced by relevant policies and is speculative.

The empirical results of this research show that the quality of green technology innovation is relatively important in determining the risk of stock market crashes. Following existing research (Kim and Zhang, 2016; Kim et al., 2011a, 2011b), we use the degree of negative stock weekly returns (NCSKEW) and the weekly return volatility ratio (DUVOL) to measure the risk of a stock price collapse.

First, the research sample includes 5,777 observation data points of 1,483 Chinese listed companies from 2008 to 2018. In the empirical analysis, we use quantitative and qualitative indicators of green technology innovation to estimate its impact on stock price crashes. The basic regression and robustness test results show that the quantity index of green technology innovation is not correlated with stock price crashes, while the quality index is negatively correlated with stock price crashes. To solve the endogeneity problem, we used the method of external policy impact on green technology innovation to conduct a difference in difference (DID) test. In 2013, the central government of China officially launched a pilot policy for the paid use and trading of carbon emission rights to provide conditions for the development of quasi-natural experiments. We used the external impact of new external policies on green technology innovation to eliminate the endogenous bias in estimating the impact of green technology innovation on the risk of stock crashes. The empirical results of the DID method show, reliably, that the quality of green technological innovation has a significant negative impact on the risk of future stock crashes.

Second, we analyzed the impact of China's unique corporate ownership on the stock price crash and found that the quality of green technology innovation has a difference in the impact of the stock price crash between state-owned enterprises and nonstate-owned enterprises. In nonstate-owned enterprises, the quality of corporate green technological innovation is negatively correlated with the risk of a stock price collapse; among state-owned enterprises, the quality of green technological innovation and the risk of a stock price collapse are positive and not significant. We believe that the reason for this phenomenon is related to the political connections of state-owned enterprises.

Finally, this study analyzes the internal mechanism by which green technology innovation affects the risk of a stock price collapse. The empirical results show that the quality of green technology innovation is positively correlated with a company’s negative news reports. First, the improvement of the quality of green technology innovation suggests there has been an improvement of the company's innovation patents, which itself reflects the company's innovation strength and influence, increases the company's information exposure, and reduces the asymmetry of information circulation. Second, the improvement of the quality of green technology innovation reflects the excellent moral example of the company's senior management and strengthens the disclosure of negative information. Third, the improvement of the quality of green technology innovation has strengthened contacts with surrounding suppliers and partners, making it more difficult to hide company information and increasing the disclosure of bad news.

Our research has contributed to the existing literature in at least six areas. First, although the existing studies have studied the asymmetric risk of corporate social responsibility(Kim et al., 2014; Zhang, et al., 2016), however, there is no research on the asymmetric risk from the perspective of quality and quantity of green technology innovation, we studies the relationship between the quantity and quality of green technology innovation and the risk of a stock price collapse, extends the theory of corporate social responsibility to the quantity and quality of green technology innovation, broadens the limited focus of previous studies that mainly analyze the impact of corporate social responsibility (Freeman et al., 1991; Stankwick et al., 1998; Simpson et al., 2002; Nelling and Webb, 2009; Barnes & Rubin, 2010) and green technology innovation(Du et al., 2015) from the income perspective introduces asymmetric risk, analyzes the economic consequences of green technology innovation from the perspective of stock price collapses, and further analyzes the relationship between interest-related behaviors and market consequences.

Second, it provides a new literature supplement to the research on the influencing factors of stock price collapse risk. In recent years, according to the management information-hiding theory proposed by Jin and Myers (2006), scholars have conducted research on the impact of corporate financial information transparency, tax avoidance, executive compensation, accounting conservatism and institutional investors on the risk of stock crashes. (Huttonetal et al., 2009; Kim et al., 2011a; Kim et al., 2011b; Kim and Zhang, 2015; Colombelli et al., 2020). The above research did not consider nonfinancial factors of the company. The company's green technological innovation is an important asset of the company, we find that the quantity and quality behavior of enterprise green technology innovation reflects the value orientation and ethical behavior of management, affects the time window of information disclosure, and is finally reflected in the index of stock price collapse.

Third, the use of the Differences-in-differences method (DID) solves the endogeneity problem of the main causal relationship in this study. Fourth, it explores the impact of corporate ownership with Chinese characteristics on the relationship between the quality of green technology innovation and the risk of a stock price collapse.

Fifth, it explores the internal mechanism between the quality of green technology innovation and the impact of stock price crash risks from the perspective of media reports, it also analyzes the difference impact of the quality of green technology innovation and the risk of stock price collapse in China's state-owned enterprises and private enterprises.

Sixth, this paper has certain practical value, for enterprise owners, this paper can further clarify whether the corporate green technology innovation strategy should be quantitative orientation or quality orientation. For policy makers, they can further make targeted use of the financial system, adjust the corresponding information disclosure regulations from the perspective of stock price risk, and pay attention to the disclosure of the quantity and quality indicators of green technology innovation. For other stakeholders, the quality of green technology innovation should be regarded as an important index to avoid investment risk, and the attention of the quantity of green technology innovation should be appropriately reduced.

At the same time as our research, Zaman (2021) used the data of U.S. companies and found that eco-innovation can inhibit the risk of stock price crashes. The main differences between our studies are as follows. First, the main variable of their research is the third-party score of the Thomson Reuters eco-innovation score and their green technology innovation evaluation index is single and one-sided, while we also analyzed the quantitative index and quality index of green innovation patents. The quantity and quality of green technological innovation are important indicators for paying attention to green technological innovation activities. Quantitative indicators represent the overall intensity of enterprise technological innovation activities. However, due to the high complexity of innovation activities, the quantitative indicators of technological innovation may be misunderstood by relevant stakeholders, because it does not mean that innovation can bring substantive innovation achievements and benefits. In contrast, the innovation quality index reflects the overall quality and innovation performance of enterprise green technology innovation, and expresses whether the enterprise's innovation has produced substantive high-level results. From the quantity and quality of patents, we find a differential impact on the stock price crash. Second, we study the differential impact of the equity situation of government participation in corporate governance with Chinese characteristics on green technology innovation and stock price crashes. Third, our research focuses on the intermediary path, mainly from the perspective of media reports. Fourth, we use regression to eliminate endogenous effects, which is more effective at dealing with endogeneity problems. Fourth, we have different theoretical frameworks and hypothesis deductions, we believe that enterprises' green technology innovation activities have value oriented view and non value oriented tool view, which may have a different impact on the stock price collapse.

## 2 | RELATED LITERATURE AND HYPOTHESIS DEVELOPMENT

### 2.1 RELATED LITERATURE

#### 2.1.1 Risk of a stock price crash

Stock crashes or plunges are an important issue to study since the 2008 financial crisis. In recent years, scholars have reached a consensus on information hidden by management. The management information-concealment hypothesis posits that the withholding of information by management causes a risk of a stock price collapse. The reasons for this behavior include the management's return on their own interests (stock options, etc.), a good career history, the company's future development plans and political factors. In information disclosure, the disclosure of good news and bad news is often inconsistent with the real time series that requires disclosure. If the information is disclosed according to the real time series and the distribution of related news is symmetric, then the stock return should also be symmetric (Kothari et al., 2009). However, related studies have shown that the distribution of information disclosure by managers is not strictly symmetrical, and management has a tendency to delay the disclosure of bad news, that is, to disclose good news in advance and delay the disclosure of bad news (Pastena and Ronen, 1979; Kothari et al., 2009). Bad news continues to accumulate within the time window. Because the company's stock price is stable, bad news needs to be in a stable interval. Once the accumulated bad news exceeds the upper limit of the interval, it will be released in a concentrated manner, causing the company's stock price to collapse (Jin & Myers, 2006; Hutton et al., 2009).

There are many empirical studies on the risk of stock crashes. Chen et al. (2001) found that the deviation of recent stock trading volume from its usual trend and past returns is an important factor. Jin and Myers (2006) found that stocks with higher information opacity are more likely to collapse. Hutton et al. (2009) and Kim and Zhang (2014) found a positive correlation between the opacity of financial reporting and the risk of stock market crashes.

On the basis of the above studies, researchers have analyzed many factors affecting stock price crashes, which can be divided into four categories. The first category is the influence of executive characteristics, such as executive stock options (Kim et al., 2011b), executive duality (Chen et al., 2015), executive overconfidence (Kim et al., 2016), senior management age (Andreou et al., 2017) and executive power (Mamun et al., 2020). The second category is the company's internal policy impact, such as corporate tax planning (Kim et al., 2011a), overtime allowance (Xu et al., 2014), social responsibility (Kim et al., 2014), accounting continuity and robustness (Kim and Zhang, 2016), corporate governance structure (Andreou et al., 2016), corporate donation behavior (Zhang et al., 2016) and acquisition protection system (Bhargava et al., 2017). The third category is the influence of the company's external informal institutions, including religious culture (Callen and Fang, 2015b), the influence of the Communist Party's political connection (Li and Chan, 2016), the degree of political connection (Luo et al., 2016; Zhang et al., 2017) and the degree of social trust (Li et al., 2017). The fourth category is the influence of the company's external formal institutions, such as institutional investors (Callen and Fang, 2013; An and Zhang, 2013), international financial analysts (Xu et al., 2013), and the impact of the implementation of international reporting standards (Defond et al. al., 2015) and short-term interests (Callen and Fang, 2015a; Ni and Zhu, 2016).

### 2.1.2 Green technology innovation

What impact will green technological innovation have on the financial behavior and performance of microenterprises? (1) Weche et al. (2019) focused on the impact of green technology innovation on business investment in the field of corporate investment decision-making and found that green technology investment crowded out other business activities. (2) In the financial field, Colombelli et al. (2020) included relevant research on the market value of green innovation for companies and found that green technology innovation can affect the value of corporate cash flow and corporate intellectual capital and then affect the value of the company in the capital market; Chuang et al. (2015) found the negative correlation between green disguise as a speculative green strategy of companies and stock returns. Haigh et al. (2004) and others studied the relationship between socially responsible investment and capital markets. Lewis et al. (2010) studied the development status and bottlenecks of green sustainable investment funds. (3) There are also many studies on the impact of green technology innovation on the productivity and competitiveness of enterprises, including the rate of return of green technology innovation on manufacturing green innovation (Marin et al., 2016), the energy green technology investment effect evaluation (Stucki et al., 2019), the impact of green IT capital (Chuang et al., 2015), the impact of green innovation on competitiveness (Chen et al., 2006), the impact of innovation as a social performance on financial performance (Hull et al., 2008), the impact of green technology innovation on financial performance (Leyva-de et al., 2019; Xie et al., 2019), the impact of green technology innovation on market performance (Pujari et al., 2006), the impact of environmental design (Tien et al., 2005), the impact of green business strategies (Leonidou et al., 2017) and the impact of new environmental products (Pujari et al., 2003).

There are few studies that directly analyze the relationship between green technology innovation and stock price crash risk; at the same time, green technological innovation is closely related to corporate social responsibility. The research study on the impact of social responsibility on stock price collapse risk mainly uses agency theory and information theory to analyze the participants' involvement in the relationship between green technology innovation and stock price crashes, including the behavior, motives and possible outcomes of each participant. There are two main points of view:

1. Management self-interest theory. The agent theory is the theoretical basis of the self-interest view of corporate social responsibility management, the agent will make decisions according to the maximization of his own interests rather than the maximization of the overall interests of the society. Scholars such as Friedman (1970) believe that the essence of corporate social responsibility activities is to serve the interests of senior executives rather than the interests of shareholders. The risk-return model of enterprises engaged in related activities is beneficial to the senior executive but not to the shareholders. Therefore, the purpose of corporate social responsibility activities is for senior executives, which only focuses on the short-term behavioral effects of corporate social responsibility, so social responsibility itself is the agency cost. Some studies believe that corporate social responsibility increases the company's additional costs and risks and is a type of damage to the competitiveness of the company (Aupperle et al., 1985; McWilliams and Siegel, 2000). Friedman denounced the theory of corporate social responsibility as an undermining of liberalism and a deviation from the optimal trajectory of corporate development, that is, thinking that the only social responsibility of a company is to maximize corporate profits according to the rules of the game.

2. Shareholderism. Resource dependence theory and stakeholder theory constitute the theoretical basis of corporate social responsibility shareholderism. The theory has also experienced continuous development, the early stage shareholderism only focused on shareholder, and the later stage shareholderism included the attention to stakeholders, the theory points out that enterprises can be regarded as a collection of relationships with shareholders, employees, suppliers, customers, consumers, communities, governments, society and other stakeholders. Resource dependence theory emphasizes that the survival of an organization needs to draw resources from its surrounding environment, and the construction of its competitive advantage depends on the core resources mastered by stakeholders. According to the theory of shareholderism, corporate social responsibility reflects the interests of shareholders and stakeholders, not the interests of management. It is believed that Shareholderism recognizes long-term effects of social responsibility of corporate social responsibility. In essence, corporate social responsibility is an asset that maintains the foundation of a relationship. Through socially responsible investment, it can attract responsible consumers (Hillman and Keim, 2001), obtain resources from socially responsible investors (Kapstein, 2001), improve financing availability (Ioannou and Serafeim, 2015), or help the company recover from financial adversity (Choi and Wand, 2009). In addition, it protects and improves the company's reputation (Fombrun and Shanley, 1990; Fombrun, 2005; Freeman et al., 2007).

Because green technology innovation naturally has the characteristics of public goods with dual externalities of environment and innovation, agents usually have no motivation to provide social optimal output. Therefore, the provision of green technology innovation products is closely related to the compliance of the management's own moral standards. This is also one of the reasons why we choose the agency theory, the information theory, the management self-interest theory and shareholder theory (stakeholders theory) as the theoretical support.

## 2.2 HYPOTHESIS DEVELOPMENT

The biggest difference between management self-interest theory and shareholderism lies in whether the behavior of management is for the interests of management itself or for the overall interests of shareholders and stakeholders, which affects the nature of corporate social responsibility behavior. Under the management self-interest theory, corporate social responsibility behavior may be instrumental(tool), that is, corporate social responsibility is only to meet the management itself, so corporate social responsibility behavior may have the element of deception, which partly reflects the unethical behavior of the management. Under the shareholderism, the purpose of corporate social responsibility is for the long-term development of the company, which partly reflects the noble moral sentiment of the management. The moral and unethical behavior of the management itself reflects whether it hides the company's negative information, which affects the time window of the company's information disclosure and the risk of stock price collapse.

The two opposing theories of management self-interest theory and shareholderism are based on differing views of corporate social responsibility. This study puts forward two opposing hypotheses in deducing the relationship between corporate social responsibility and the risk of a stock price collapse, namely, the value hypothesis and the instrument(tool) hypothesis.

1. The value hypothesis holds that a company’s fulfillment of its social responsibility is oriented more toward shareholderism and long-term value. In this case, it is expected that the better the corporate social responsibility, the more investors and shareholders will recognize its value and the longer they are willing to hold the company’s shares and the lower the risk of a crash. The reasons for this dynamic are as follows. (1) From the perspective of agency theory, if a company truly pays attention to the interests of shareholders in fulfilling its social responsibilities, its good social responsibility performance reflects higher management's moral standards and there is a lower probability that management will engage in self-interested behavior by hiding information. (2) From the perspective of information theory, if corporate social responsibility investment is to make decisions for the long-term development of the company, then management will pay attention to the bond relationship with investors, consumers, suppliers and other stakeholders. Management will provide reliable internal information that meets the needs of stakeholders, reduces the degree of information asymmetry between internal and external stakeholders, enhances the ability of stakeholders to supervise the enterprise, and improves the likelihood that any information hidden by management will be discovered. Gelb and Strawser (2001) found that corporate social responsibility is positively related to financial information disclosure. Kim et al. (2012) found that corporate social responsibility is negatively related to earnings management. Jin and Myers (2006), Huton et al. (2009) and Francis et al. (2016) found that corporate information transparency is negatively related to the risk of collapse.

Green technological innovation activities can also be considered a special form of corporate social responsibility. From the perspective of the value hypothesis, if green technology innovation meets the abovementioned agency theory and information theory definitions, it is believed that green technological innovation exemplifies management diligence and responsibility, strengthens the external environment's supervision of management, reduces the degree of information asymmetry between the internal and external enterprises, and ultimately reduces the probability of management information-hiding behavior, making the company's information disclosure time window as consistent as possible with the real information disclosure time window, and ultimately, reducing the risk of a stock price collapse.

Therefore, under the value hypothesis, based on the quantity and quality indicators of green technology innovation, this article proposes the following hypotheses:

H1a: If the value hypothesis is established, the relationship between the quantity of green technological innovations and the risk of a stock price collapse will be significantly negative.

H2a: If the value hypothesis is established, the relationship between the quality of green technology innovation and the risk of a stock price collapse will be significantly negative.

2. The tool hypothesis is based on management self-interest theory. From the perspective of agency theory, management may use the packaging of corporate social responsibility to enhance its professional reputation and personal interests, and this self-interested disguise is at the expense of shareholders' interests. Hemingway and Madagan (2004) argue that the company's motivation for social responsibility is to cover up management's unethical behavior and negative news. Enron is a typical representative of this perspective. This company, which caused a sensation in the world because of its accounting fraud, had won many social responsibility awards before the incident and was once regarded as a model of social responsibility. From the perspective of information theory, many documents have found that corporate social responsibility and earnings management are positively correlated (Petrovits, 2006), which in turn distorts financial information disclosure and prompts management to conceal information. If corporate social responsibility essentially embodies management self-interest theory, then corporate social responsibility and management information concealment have the same motivation and direction, which reflects management's interest in pursuing power for personal gain. Therefore, the social responsibility of enterprises and the risk of a stock price collapse should be positively related. Furthermore, management's use of corporate social responsibility as a tool to conceal financial statement information will worsen the company's transparency, strengthen the management's information-concealment behavior, and exacerbate the risk of a stock price collapse.

From the tool hypothesis, so if green technology innovation meets the above definitions of agency theory, and information theory, it is believed that green technology innovation embodies the self-interested behavior of management, and the purpose is to maximize benefits and avoid external supervision, and improve the probability of information-concealment behavior of management of the company, and further expands the inconsistency between the company's information disclosure time window and the real information disclosure time window, and ultimately increases the risk of a stock price collapse.

Therefore, under the tool hypothesis, based on the quantity and quality indicators of green technological innovation, this paper proposes the following hypotheses:

H1b: If the tool hypothesis is established, the relationship between the quantity of green technological innovation and the risk of a stock price collapse will be significantly positive.

H2b: If the tool hypothesis is established, the relationship between the quality of green technology innovation and the risk of a stock price collapse will be significantly positive.

3. Due to the different agency mechanisms and codes of conduct between the relevant management of state-owned enterprises and private enterprises, we believe that this will affect the information hiding behavior mode of management, and affect the relationship between green technology innovation and stock price collapse. Compared with other countries, the Chinese enterprise ownership system has distinct characteristics; in particular, the state has a certain degree of ownership and control over some enterprises. State-owned enterprises enjoy various forms of political and financial support from the government (Zhang et al., 2010; Ye and Zhang, 2011; Zhang et al., 2014). The government may provide assistance to state-owned enterprises in financial difficulties (Wang et al., 2008). Therefore, the profit volatility of state-owned enterprises is less than that of nonstate-owned enterprises, and the risk of a stock price collapse should be relatively small. At the same time, the development of enterprises is closely related to government policies and resource tilt. Luo et al. (2016) found that the stock crash risk of politically affiliated companies is lower than that of their peers. They believe that political relations may prompt these companies to release bad news on time, thereby reducing their risk of a stock price collapse. In contrast, nonstate-owned enterprise boards have fewer people with political connections, so they are more likely to hide bad news.

We believe that if there is a relationship between green technological innovation and the risk of stock crashes, there should be a difference in the risk between nonstate-owned and state-owned enterprises. Therefore, we propose the following hypotheses:

H3: The impact of green technological innovation quality promotion behavior on the risk of stock collapse of nonstate-owned enterprises is different from that of state-owned enterprises.

4. Intermediary mechanism

As the share price collapse is mainly due to the concealment of negative information by the management(Jin & Myers, 2006; Hutton et al., 2009), the relevant media reports, especially the number of negative media reports, may play an important intermediary role between green innovation and stock price collapse. Media reports are an effective way for external stakeholders to understand the true situation of a company. Regardless of whether the purpose of corporate information disclosure is "self-interest," companies tend to enhance their influence and status through the media when conducting relevant social responsibility activities. To meet public needs and promote its own development, the media will also pay attention to corporate social responsibility activities in a timely manner (Dyck et al., 2004). Corporate social responsibility actions are highly relevant to media reports for the following reasons. First, from the standpoint of media responsibility, the mission of news media is to encourage corporate green responsible behaviors, supervise related corporate behaviors, expose false social responsibility behaviors that are packaged or disguised, and play the role of guiding and supervising public opinion. Second, as the public's awareness of environmental protection has increased, stakeholders have become increasingly concerned about corporate social responsibility. As a relatively autonomous third party, the media's independent status better meets the needs of stakeholders for information.

Therefore, theoretically, the corporate social responsibility index is positively correlated with the number of media reports. In this study, corporate green technology innovation behavior itself also has certain social responsibility attributes. In the analysis, since a stock price collapse is mainly related to insufficient and the lack of timely reports of negative corporate information, we mainly need to pay attention to the relationship between green technological innovation and the number of negative media reports. In theory, if the "value hypothesis" of green technology innovation is established, it reflects the noble moral sentiment of the management, it also reflects the high degree of interaction with external participants in green technology innovation activities, which will reduce the concealment of negative information, then the company's green technology innovation should positively correlated with the company's negative news disclosure; conversely, if the "tool hypothesis" is established, the improvement of the company's green technology innovation reflects a company's information-hiding behavior, in which case corporate green technological innovation is negatively correlated with corporate negative news disclosure. The actual impact still needs to be further tested and analyzed through empirical research.

To this end, the following hypotheses are made:

H4a: Green technology innovation quality promotion behavior is positively correlated with the number of negative media reports.

H4b: Green technological innovation quality promotion behavior is negatively correlated with the number of negative media reports.

Fang and Peress (2009) believe that the media, as an effective information intermediary, has the functions of information supervision and transmission. Peress (2014), based on newspapers' collective workers' strike behavior, found a significant 12% drop in market transaction volume and a 7% reduction in stock price volatility, indicating that newspapers, as an information medium, can significantly affect stock market volatility. As a relatively independent third-party role, news media has a strong public opinion-oriented function and acts as an external mechanism for the supervision of listed companies. Their behavior will also affect the fluctuation of the company's stock price. If a listed company is inclined to shareholderize, it will take the initiative to spread negative corporate news in a timely manner through the media to reduce the risk of a stock price collapse. Of course, it cannot be ruled out that management may confuse competitors by actively releasing false negative news. In this way, the impact of negative information reports of listed companies and stock price crashes may be uncertain. Therefore, the impact of negative news reports of listed companies on stock price crashes needs to be further tested empirically.

To this end, the following hypotheses are made:

H5a: The number of negative media reports by listed companies is negatively correlated with the risk of a stock price collapse.

H5b: The number of negative media reports of listed companies is positively correlated with the risk of a stock price collapse.

## 3| SAMPLE DESCRIPTION AND VARIABLE MEASUREMENT

### 3.1 Data source and sample selection

This study collects relevant data from the databases of the China Stock Market & Accounting Research Database (CSMAR) and Chinese Research Data Services (CNRDS) and uses the main financial indicators and data listed on the Chinese A-share market from 2008 to 2018 as the initial sample. The classification standard of green technology innovation patents is distinguished according to the green patent classification number standard published by the World Intellectual Property Office. In the data collection process, we tried our best to ensure that the sample size was maximized. According to the research of Xu (2014) and others, first, we exclude financial companies because their accounting reporting rules and capital structures are different from ordinary companies. Second, we excluded companies with stock return data less than 30 trading weeks within a year. Finally, we eliminated the insufficient and missing data of related variables， . The final sample included 5,777 company annual observations, representing 1,483 unique companies.

Table 1 lists sample distribution.

**Table 1 Sample distribution**

| Year | n | %_Total |
|---|---|---|
| 2008 | 74 | 1.28% |
| 2009 | 118 | 2.04% |
| 2010 | 190 | 3.29% |
| 2011 | 283 | 4.90% |
| 2012 | 460 | 7.96% |
| 2013 | 499 | 8.64% |
| 2014 | 574 | 9.94% |
| 2015 | 659 | 11.41% |
| 2016 | 813 | 14.07% |
| 2017 | 958 | 16.58% |
| 2018 | 1,149 | 19.89% |
| total | 5777 | |

### 3.2| Variable description

#### 3.2.1 Green Innovation

1. Quantity index $NGInnovation1$

The number of patents have the advantage of being a good indicator of innovation(Comanor and Scherer, 1969; Griliches,1990; Hagedoorn and Cloodt, 2003).Taking into account that the company's green patent information disclosure may have been disclosed in advance during the research and development process, which may have an impact on the capital market, we use the total number of green invention patents and utility model patent applications of listed companies as a proxy variable for the number of green technology innovation indicators $NGInnovation1$.

In addition, in the robustness test part, consider the comparability of the number of green technology innovation patents of companies of different sizes, we use the total number of green invention patents and utility model patent applications of listed companies divided by the natural logarithm of the size of the company in the year as a proxy variable for the number of green technology innovation indicators $NGInnovation2$.

2. Quality Index $QGInnovation$

Patents that continuously generate citations are viewed as higher-quality patented innovations (Hall et al.,2005). To measure the impact of green patents, we pay attention to the index of the number of citations of green patents as a key indicator for measuring influence. Because the impact of green patents may take several years to officially decline, considering the role of green patents, the total number of green patent applications and the number of cumulative green patent citations, the establishment of indicators to measure the quality of the company's green patents is as follows:

$$\text{Green patent quality}_{i,t} = \frac{\sum_{2008}^{t} \text{Total number of green patents cited}_{i,t}}{\sum_{2008}^{t} \text{Number of green patent applications}_{i,t}}$$

### 3.2.2 Stock price crash risk

Drawing lessons from previous methods (Kim and Zhang, 2016; Kim et al., 2011a, 2011b), we constructed two common stock crash risk measures: NCSKEW and DUVOL. For these two indicators, we first use the weekly stock return rate to estimate the time series model for each company and year. The residual items that cannot be explained by the model are the focus of our attention because this is closely related to the risk of a stock price collapse. As shown in formula (1):

$$r_{i,t} = \alpha_i + \beta_1 r_{m,t-2} + \beta_2 r_{m,t-1} + \beta_3 r_{m,t} + \beta_4 r_{m,t+1} + \beta_5 r_{m,t+2} + \varepsilon_{i,t} \qquad (1)$$

where $r_{i,t}$ is the return on stock i in week t, and $r_{m,t}$ is the value-weighted market return in week t.

According to the estimation of formula (1), we calculate the natural logarithm of the sum of the residual term and 1, $w_{i,t} = \ln(1 + \varepsilon_{i,t})$.

The first indicator to measure the risk of a stock market crash is the negative conditional skewness (NCSKEW) of a company's specific weekly rate of return in a fiscal year. We calculate the NCSKEW of each company i in year t, as shown in the following formula:

$$NCSKEW_{i,t} = -\left[n(n-1)^{3/2} \sum W_{i,t}^3\right] / \left[(n-1)(n-2) \left(\sum W_{i,t}^2\right)^{3/2}\right]$$

where n is the number of trading weeks for stock i in year t. The higher the value of NCSKEW is, the higher the risk of stock crashes. The second indicator is the up and down volatility (DUVOL), which is calculated as follows:

$$\text{DUVOL}_{i,t} = \log\left\{\left[(n_u - 1) \sum_{Down} W_{i,t}^2\right] / \left[(n_d - 1) \sum_{Up} W_{i,t}^2\right]\right\}$$

where $n_u$ and $n_d$ are the number of up and down weeks, respectively. The higher the values of NCSKEW and DUVOL, the greater the risk of a collapse in stock prices.

### 3.3| The benchmark model

To examine the relation between green innovation and stock crash risk, we estimate the following equations (2)-(3):

(2)

$$NCSKEW_{i,t+1}\left(DUVOL_{i,t+1}\right)=\beta_0+\beta_1 NGInnovation_{i,t}+\sum\beta_j\left(Control\ variables\right)_{i,t}+\sum Firm_{i,t}+\sum Year+\varepsilon_{i,t}$$

(3)

$$NCSKEW_{i,t+1}\left(DUVOL_{i,t+1}\right)=\beta_0+\beta_1 QGInnovation_{i,t}+\sum\beta_j\left(Control\ variables\right)_{i,t}+\sum Firm_{i,t}+\sum Year+\varepsilon_{i,t}$$

$NCSKEW_{i,t+1}$ and $DUVOL_{i,t+1}$ refer to the crash risk of stock i in year t. The dependent variable is measured in year t+1, while the independent variables are measured in year t. Year and industry fixed effects are included to control for time- and industry-invariant factors.

The variables of interest, $NGInnovation_{i,t}$ and $QGInnovation_{i,t}$, measure the quantity and quality of green innovation. A larger $\beta_1$ represents a greater impact of green innovation.

### 3.4| control variables

$NCSKEW_{i,t}\left(DUVOL_{i,t}\right)$ is the negative conditional skewness of firm-specific weekly returns of stock i in year t, to account for the potential serial correlation of $NCSKEW_{i,t}\left(DUVOL_{i,t}\right)$, we control this variable (Kim et al., 2011b). $Sigma_{i,t}$ is the standard deviation of firm-specific weekly returns of stock i over fiscal year period t, since more volatile stocks are more prone to collapse, we control this variable (Kim et al.,2014 and Callen and Fang, 2015a). The research of Chen et al. (2001) shows that the historical return of stock price will affect the risk of stock collapse, we add control variable $Ret_{i,t}$, $Ret_{i,t}$ is the mean of firm-specific weekly returns of stock i over fiscal year t. Chen et al. (2001) show that trading volume, a proxy for the intensity of differences of opinion among investors, is a predictor of stock price crash risk, We add control variable $Oturnover_{i,t}$, $Oturnover_{i,t}$ is the detrended average monthly stock turnover of stock i in year t, calculated as the average monthly share turnover of stock i in year t minus the average monthly share turnover of stock i in year t-1.

$ABACC_{i,t}$ is the absolute value of the estimated residuals (accruals) from the adjusted Jones model (Dechow et al., 1995) of stock i in year t, relevant studies have found that corporate earnings management activities are related to the risk of stock price collapse(Hutton et al., 2009).

Relevant studies show that the company's fundamental factors will affect the risk of stock price collapse. Based on this, add the following control variables: $Size_{i,t}$ is the natural logarithm of the book value of total assets of stock i at the end of year t, relevant studies show that enterprise scale has an impact on stock price collapse (Harvey and Siddique, 2000; Chen et al., 2001). $BM_{i,t}$ is the market-to-book ratio, measured as the market value of equity divided by the book value of equity of stock i in year t(Kim et al.,2014). $Lev_{i,t}$ is firm leverage, calculated by the book value of total debt divided by the book value of total assets of stock i in year t(Shahab et al.,2014). $Roa_{i,t}$ is the return on assets, computed as net profit divided by the book value of total assets of stock i in year t(Shahab et al.,2014).

Relevant studies show that there is a close relationship between the company's ownership structure and the stock price collapse, so we add the following control variables： $Tophold_{i,t}$ refers to the shareholding ratio of the largest shareholder of company I in year t(Yuan et al,2016). $Power_{i,t}$ refers to the management power dummy variable of stock i in year t. When the chairman and CEO hold two posts concurrently, it is equal to 1; otherwise, it is 0(Shahab et al.,2014). $Maghold_{i,t}$ The management shareholding ratio is a dummy variable of stock i in year t. When the management's shareholding ratio is greater than the year and industry median, it is 1(Li et al.,2021); otherwise, it is 0. $Balance_{i,t}$ refers to the equity balance dummy variable of stock i in year t. When the equity balance degree of a company is greater than the annual and industry median, it is 1; otherwise, it is 0(Andreou et al.,2016). $Inshold_{i,t}$ refers to institutional investors holding dummy variables of stock i in year t. When the proportion of institutional investors is greater than the annual and industry median, it is 1; otherwise, it is 0. (Yuan et al,2016)

## 4| EMPIRICAL RESULTS

### 4.1| Descriptive statistics and correlation analysis

Table 2-1 reports descriptive statistics of stock crash risk, green innovation, and control variables for the 2008–2018 sample. Table 2-2 reports the Pearson correlation coefficients between the explanatory variables. the correlation coefficients between independent variables and control variables were less than 0.5, So there is no evidence of severe multicollinearity among the explanatory variables.

**Table 2-1 Descriptive statistics results**

| | (1) | (2) | (3) | (4) | (5) |
|---|---|---|---|---|---|
| VARIABLES | N | mean | sd | min | max |
| | | | | | |
| $NCSKEW_{i,t}$ | 5,777 | -0.259 | 0.739 | -3.878 | 4.166 |
| $DUVOL_{i,t}$ | 5,777 | -0.175 | 0.495 | -2.046 | 2.287 |
| $Ret_{i,t}$ | 5,777 | 0.001 | 0.010 | -0.045 | 0.0707 |
| $Sigma_{i,t}$ | 5,777 | 0.063 | 0.025 | 0.015 | 0.252 |
| $ABACC_{i,t}$ | 5,777 | 0.053 | 0.054 | 2.36e-05 | 0.578 |
| $Tophold_{i,t}$ | 5,777 | 0.345 | 0.153 | 0.003 | 0.852 |
| $Power_{i,t}$ | 5,777 | 0.229 | 0.420 | 0 | 1 |
| $Maghold_{i,t}$ | 5,777 | 0.434 | 0.496 | 0 | 1 |
| $Balance_{i,t}$ | 5,777 | 0.468 | 0.499 | 0 | 1 |
| $Inshold_{i,t}$ | 5,777 | 0.610 | 0.488 | 0 | 1 |
| $Oturnover_{i,t}$ | 5,777 | -0.104 | 0.425 | -3.588 | 3.285 |
| $Size_{i,t}$ | 5,777 | 22.73 | 1.368 | 19.485 | 28.253 |
| $BM_{i,t}$ | 5,777 | 0.645 | 0.245 | 0.052 | 1.363 |
| $Lev_{i,t}$ | 5,777 | 0.472 | 0.195 | 0.022 | 1.352 |
| $Roa_{i,t}$ | 5,777 | 0.036 | 0.069 | -1.909 | 0.400 |
| $NGInnovation1_{i,t}$ | 5,777 | 19.44 | 65.35 | 1.000 | 1,566 |
| $NGInnovation2_{i,t}$ | 5,777 | 0.811 | 2.551 | 0.037 | 58.760 |
| $QGInnovation_{i,t}$ | 5,777 | 0.988 | 1.502 | 0.000 | 60 |
| $Soe_{i,t}$ | 5,777 | 0.430 | 0.495 | 0 | 1 |
| $ONMEDIA_{i,t}$ | 5,775 | 33.766 | 73.180 | 0 | 2309 |
| $ANMEDIA_{i,t}$ | 5,775 | 132.510 | 344.995 | 0 | 12921 |

**Table 2-2 The correlation matrix of the explanatory variables.**

| Variables | (1) | (2) | (3) | (4) | (5) | (6) | (7) | (8) |
|---|---|---|---|---|---|---|---|---|
| $NGInnovation2_{i,t}$ | 1.00 | | | | | | | |
| $QGInnovation_{i,t}$ | -0.04a | 1.00 | | | | | | |
| $NGInnovation1_{i,t}$ | 1.00a | -0.03b | 1.00 | | | | | |
| $Ret_{i,t}$ | -0.01 | 0.05a | -0.01 | 1.00 | | | | |
| $Sigma_{i,t}$ | -0.08a | 0.03b | -0.08a | 0.50a | 1.00 | | | |
| $ABACC_{i,t}$ | -0.03b | 0.07a | -0.03b | 0.01 | 0.10a | 1.00 | | |
| $Tophold_{i,t}$ | 0.03b | -0.07a | 0.04a | 0.03b | -0.09a | -0.04a | 1.00 | |
| $Power_{i,t}$ | 0.00 | 0.02 | 0.00 | 0.01 | 0.06a | 0.06a | -0.07a | 1.00 |
| $Maghold_{i,t}$ | -0.06a | 0.01 | -0.06a | 0.03b | 0.10a | 0.09a | -0.26a | 0.21a |
| $Balance_{i,t}$ | -0.01 | 0.03b | -0.01 | 0.00 | 0.03b | 0.01 | -0.42a | 0.05a |
| $Inshold_{i,t}$ | 0.08a | -0.02 | 0.08a | 0.03b | -0.11a | -0.07a | 0.33a | -0.16a |
| $Oturnover_{i,t}$ | 0.01 | 0.04a | 0.01 | 0.40a | 0.34a | -0.03b | -0.02c | -0.05a |
| $Size_{i,t}$ | 0.29a | -0.05a | 0.30a | -0.08a | -0.27a | -0.10a | 0.28a | -0.18a |
| $BM_{i,t}$ | 0.10a | -0.12a | 0.10a | -0.42a | -0.35a | -0.11a | 0.18a | -0.14a |
| $Lev_{i,t}$ | 0.13a | -0.08a | 0.13a | -0.02 | -0.07a | 0.04a | 0.14a | -0.15a |
| $Soe_{i,t}$ | 0.05a | -0.07a | 0.06a | -0.00 | -0.12a | -0.09a | 0.31a | -0.30a |

| Variables | (9) | (10) | (11) | (12) | (13) | (14) | (15) | |
|---|---|---|---|---|---|---|---|---|
| $NGInnovation2_{i,t}$ | | | | | | | | |
| $QGInnovation_{i,t}$ | | | | | | | | |
| $NGInnovation1_{i,t}$ | | | | | | | | |
| $Ret_{i,t}$ | | | | | | | | |
| $Sigma_{i,t}$ | | | | | | | | |
| $ABACC_{i,t}$ | | | | | | | | |
| $Tophold_{i,t}$ | | | | | | | | |
| $Power_{i,t}$ | | | | | | | | |
| $Maghold_{i,t}$ | 1.00 | | | | | | | |
| $Balance_{i,t}$ | 0.25a | 1.00 | | | | | | |
| $Inshold_{i,t}$ | -0.41a | -0.15a | 1.00 | | | | | |
| $Oturnover_{i,t}$ | -0.08a | -0.03b | 0.04a | 1.00 | | | | |
| $Size_{i,t}$ | -0.28a | -0.04a | 0.27a | 0.09a | 1.00 | | | |
| $BM_{i,t}$ | -0.17a | -0.03a | 0.00 | -0.06a | 0.43a | 1.00 | | |
| $Lev_{i,t}$ | -0.23a | -0.09a | 0.13a | 0.07a | 0.41a | 0.43a | 1.00 | |
| $Soe_{i,t}$ | -0.42a | -0.23a | 0.30a | 0.09a | 0.37a | 0.27a | 0.31a | 1.00 |

This table reports pairwise Pearson correlation coefficients between the variables. a, b, and c indicate significance at the 1%, 5%, and 10% levels, respectively.

## 4.2| Multivariate analysis

### 4.2.1 Test of Hypothesis 1

According to the regression results in Table 3(1)-(2), the relationship between the quantity indicator of green technology innovation and the stock price collapse indicator NCSKEW is positive but not significant. Neither H1a nor H1b holds, which means that simply increasing the number of green technological innovations by companies does not significantly affect the stock price crash. We believe that the reason for this outcome is that there is both a "value" and "instrumental(tool)" impact of the quantity of green technological innovation. Since quantitative indicators do not reflect the quality of patents, in the full sample, some companies adopt an "instrumental(tool)" approach. Poor manufacturing and low-quality green patents have a positive impact on stock price crashes, which offsets the negative impact of high-quality companies' "value-based" research and the development of high-quality green technology innovation on stock price crashes. On the whole, the effects of the two forces cancel each other out, resulting in the insignificance of the regression measurement results. According to the regression results in Table 3(3), the regression results of the robustness test using the GMM method are listed, and the regression results of the core variables are consistent with Table 3(2). In addition, the regression results of the control variables are basically consistent with expectations.

Table 3 Quantity of green innovation on stock price crash risk

| *Model* | *FE* | | | *GMM* |
|---|---|---|---|---|
| *VARIABLES* | (1)<br>$NCSKEW_{i,t+1}$ | (2)<br>$NCSKEW_{i,t+1}$ | *VARIABLES* | (3)<br>$NCSKEW_{i,t+1}$ |
| $NCSKEW_{i,t}$ | | $-0.1278^{***}$<br>(-8.6283) | $NCSKEW_{i,t}$ | $0.0424^{*}$<br>(1.6533) |
| $NGInnovation1_{i,t}$ | 0.0003 | 0.0002 | $NGInnovation1_{i,t}$ | 0.0008* |
| | (1.4231) | (0.8221) | | (1.8670) |
| $Oturnover_{i,t}$ | | -0.0327 | $Oturnover_{i,t}$ | -0.0572 |
| | | (-0.9288) | | (-1.1941) |
| $Sigma_{i,t}$ | | -0.8002 | $Sigma_{i,t}$ | 5.3382*** |
| | | (-0.9374) | | (3.6606) |
| $Ret_{i,t}$ | | 4.4507*** | $Ret_{i,t}$ | 3.5672 |
| | | (2.6621) | | (1.4937) |
| $Size_{i,t}$ | | 0.1162*** | $Size_{i,t}$ | 0.2135** |
| | | (2.9181) | | (2.4473) |
| $BM_{i,t}$ | | -0.5971*** | $BM_{i,t}$ | -0.9184*** |
| | | (-4.8590) | | (-4.2979) |
| $Lev_{i,t}$ | | -0.5002*** | $Lev_{i,t}$ | -0.4490 |
| | | (-3.4293) | | (-1.6431) |
| $Roa_{i,t}$ | | -0.5946*** | $Roa_{i,t}$ | -0.3615 |
| | | (-2.7826) | | (-1.2988) |
| $ABACC_{i,t}$ | | 0.4021* | $ABACC_{i,t}$ | 0.5373* |
| | | (1.8436) | | (1.6980) |
| $Tophold_{i,t}$ | | 0.1527 | $Tophold_{i,t}$ | 0.2976 |
| | | (0.5888) | | (0.5512) |
| $Power_{i,t}$ | | 0.0353 | $Power_{i,t}$ | -0.0069 |
| | | (0.7799) | | (-0.0885) |
| $Maghold_{i,t}$ | | -0.0187 | $Maghold_{i,t}$ | -0.0199 |
| | | (-0.4520) | | (-0.2766) |
| $Balance_{i,t}$ | | 0.1214*** | $Balance_{i,t}$ | 0.1679** |
| | | (2.6850) | | (2.2988) |
| $Inshold_{i,t}$ | | 0.0307 | $Inshold_{i,t}$ | -0.0745 |
| | | (0.9050) | | (-1.4778) |
| *Constant* | | -2.3790*** | | |
| | | (-2.8973) | | |
| *Firm FE* | | *Yes* | *Firm FE* | *Yes* |
| *Year FE* | | *Yes* | *Year FE* | *Yes* |
| | | | *Observations* | 3,658 |
| | | | *Number of firms* | 1,044 |
| *Observations* | 5,777 | 5,777 | *AR*(1) | -11.39*** |
| | | | | (Pr > z = 0.000) |
| *Number of firms* | 1,483 | 1,483 | *AR*(2) | 0.91 |
| | | | | (Pr > z = 0.364) |
| $R^2$ | 0.049 | 0.085 | *Sargan test* | 106.84 |
| | | | | (Prob > chi2 = 0.513) |

This table reports the estimated results from the regressions of the quantity of green innovation on stock price crash risk. The dependent variable, NCKEW, is measured as of year t+1, while the independent variables are measured as of year t. For the fixed effect model, t statistic based on the robust standard error is in parentheses, and for the GMM model, z statistic based on the robust standard error is in parentheses. ***, **, and * indicate significance at the 1%, 5%, and 10% levels, respectively.

**4.2.2 Test of Hypothesis 2**

According to the regression results in Table 4 (1)-(2), the relationship between the quality index of green technology innovation and NCSKEW is negative, which is significant at the 5% confidence level. H2b is established and H2a is not established, which indicates that the risk of a stock price collapse can be reduced by long-term improvements in the quality of green innovation. We believe that the reason for this is that the quality index of green technology innovation has "value". On the one hand, it shows that, according to agency theory, the improvement of green innovation reflects management's good information disclosure habits; on the other hand, it also expresses that the time window of management information disclosure is close to the real information disclosure window, and the probability of self-interest information concealment is relatively low. Meanwhile, according to information theory, enterprises engaged in improving the quality of green technology innovation, indicating that the enterprise has established a good relationship network that is, in turn, conducive to the timely interaction between information and the outside world, reduces the blind area of information of investors, and, thus, reduces the possibility of a stock price collapse. Table 4 (3) lists the regression results of the GMM robustness test. The regression results of the core variables are consistent with those in Table 4 (2). In addition, the regression results of the control variables are basically consistent with expectations.

Table 4 Quality of green innovation on stock price crash risk

| Model | FE | | | GMM |
|---|---|---|---|---|
| | (1) | (2) | | (3) |
| VARIABLES | $NCSKEW_{i,t+1}$ | $NCSKEW_{i,t+1}$ | VARIABLES | $NCSKEW_{i,t+1}$ |
| $NCSKEW_{i,t}$ | | $-0.1278^{***}$ | $NCSKEW_{i,t}$ | $0.0424^{*}$ |
| | | (-8.6337) | | (1.6977) |
| $QGInnovation_{i,t}$ | -0.0180** | $-0.0215^{**}$ | $QGInnovation_{i,t}$ | $-0.0299^{***}$ |
| | (-1.9635) | (-2.4290) | | (-2.6017) |
| $Oturnover_{i,t}$ | | -0.0333 | $Oturnover_{i,t}$ | -0.0467 |
| | | (-0.9487) | | (-0.9820) |
| $Sigma_{i,t}$ | | -0.8648 | $Sigma_{i,t}$ | $4.0674^{***}$ |
| | | (-1.0170) | | (2.9455) |
| $Ret_{i,t}$ | | $4.4993^{***}$ | $Ret_{i,t}$ | $6.0526^{**}$ |
| | | (2.6961) | | (2.5195) |
| $Size_{i,t}$ | | $0.1204^{***}$ | $Size_{i,t}$ | $0.1753^{**}$ |
| | | (3.0618) | | (2.0687) |
| $BM_{i,t}$ | | $-0.6028^{***}$ | $BM_{i,t}$ | $-0.7796^{***}$ |
| | | (-4.9136) | | (-3.7010) |
| $Lev_{i,t}$ | | $-0.5103^{***}$ | $Lev_{i,t}$ | $-0.5360^{**}$ |
| | | (-3.5082) | | (-2.0212) |
| $Roa_{i,t}$ | | $-0.6025^{***}$ | $Roa_{i,t}$ | $-0.4681^{*}$ |
| | | (-2.8438) | | (-1.7241) |
| $ABACC_{i,t}$ | | $0.4116^{*}$ | $ABACC_{i,t}$ | $0.5284^{*}$ |
| | | (1.8919) | | (1.6741) |
| $Tophold_{i,t}$ | | 0.1712 | $Tophold_{i,t}$ | 0.3893 |
| | | (0.6614) | | (0.7183) |
| $Power_{i,t}$ | | 0.0354 | $Power_{i,t}$ | -0.0095 |
| | | (0.7841) | | (-0.1210) |
| $Maghold_{i,t}$ | | -0.0228 | $Maghold_{i,t}$ | 0.0168 |
| | | (-0.5501) | | (0.2264) |
| $Balance_{i,t}$ | | $0.1252^{***}$ | $Balance_{i,t}$ | $0.1518^{**}$ |
| | | (2.7731) | | (2.1903) |
| $Inshold_{i,t}$ | | 0.0302 | $Inshold_{i,t}$ | -0.0748 |
| | | (0.8909) | | (-1.4339) |
| Constant | -0.178*** | $-2.4594^{***}$ | | |
| | (0.000) | (-3.0310) | | |
| Firm FE | Yes | Yes | Firm FE | Yes |
| Year FE | Yes | Yes | Year FE | Yes |
| | | | Observations | 3,658 |
| | | | Number of firms | 1,044 |
| Observations | 5,777 | 5,777 | AR(1) | -11.42*** |
| | | | | (Pr> z |
| | | | | =0.000) |
| *Number of firms* | 1,483 | 1,483 | *AR*(2) | 0.98 (Pr>z = 0.326 ) |
| $R^2$ | 0.0003 | 0.086 | *Sargan test* | 107.89 (Prob>chi2= 0.485) |

This table reports the estimated results from the regressions of the quality of green innovation on stock price crash risk. The dependent variable, NCSKEW, is measured as of year t+1, while the independent variables are measured as of year t. For the fixed effect model, t statistic based on the robust standard error is in parentheses, and for the GMM model, z statistic based on the robust standard error is in parentheses. ***, **, and * indicate significance at the 1%, 5%, and 10% levels, respectively.

### 4.3| Robustness tests

First, according to table 5-1 and table 5-2, we replaced the measurement method of the quantity of green innovation, and the result is consistent with the benchmark regression. Second, to ensure the robustness of the results, we replace the result variable NCSKEW with DUVOL. According to the regression results of Tables 5-3 (1) –(2)and 6 (1)-(2), the quantitative index of green technology innovation is not related to the stock price crash, while the relationship between the quality index of green technology innovation and DUVOL is negative and true at the 1% confidence level. At the same time, according to the results of the GMM regression in Table 5-3 (3) and Table 6 (3), the measurement results of relevant core variables are consistent with the basic regression.

Table 5−1 Robustness tests of the quantity of green innovation on stock price crash risk

| *Model* | *FE* | | | |
|---|---|---|---|---|
| | (1) | (2) | (3) | (4) |
| *VARIABLES* | $NCSKEW_{i,t+1}$ | $NCSKEW_{i,t+1}$ | $DUVOL_{i,t+1}$ | $DUVOL_{i,t+1}$ |
| $NCSKEW_{i,t}$ | | -0.1278*** | | |
| | | (-8.6265) | | |
| $DUVOL_{i,t}$ | | | | -0.1228*** |
| | | | | (-8.2282) |
| $NGInnovation2_{i,t}$ | 0.0078 | 0.0046 | 0.0065 | 0.0043 |
| | (1.2737) | (0.7062) | (1.5994) | (1.0376) |
| $Oturnover_{i,t}$ | | -0.0326 | | -0.0152 |
| | | (-0.9283) | | (-0.6754) |
| $Sigma_{i,t}$ | | -0.8054 | | -0.7270 |
| | | (-0.9435) | | (-1.2676) |
| $Ret_{i,t}$ | | 4.4594*** | | 3.1361*** |
| | | (2.6680) | | (2.7430) |
| $Size_{i,t}$ | | 0.1167*** | | 0.0647** |
| | | (2.9256) | | (2.4436) |
| $BM_{i,t}$ | | -0.5975*** | | -0.3827*** |
| | | (-4.8617) | | (-4.8203) |
| $Lev_{i,t}$ | | -0.5012*** | | -0.2920*** |
| | | (-3.4364) | | (-2.9621) |
| $Roa_{i,t}$ | | -0.5953*** | | -0.3865** |
| | | (-2.7853) | | (-2.5766) |
| $ABACC_{i,t}$ | | 0.4017* | | 0.4664*** |
| | | (1.8415) | | (3.1742) |
| $Tophold_{i,t}$ | | 0.1536 | | 0.0288 |
| | | (0.5922) | | (0.1642) |
| $Power_{i,t}$ | | 0.0354 | | 0.0159 |
| | | (0.7833) | | (0.5579) |
| $Maghold_{i,t}$ | | -0.0188 | | -0.0242 |
| | | (-0.4556) | | (-0.8548) |
| $Balance_{i,t}$ | | 0.1214*** | | 0.0711** |
| | | (2.6853) | | (2.3730) |
| $Inshold_{i,t}$ | | 0.0307 | | 0.0250 |
| | | (0.9044) | | (1.0826) |
| *Constant* | | -2.3877*** | | -1.3006** |
| | | (-2.9049) | | (-2.3398) |
| *Firm FE* | *Yes* | *Yes* | *Yes* | *Yes* |
| *Year FE* | *Yes* | *Yes* | *Yes* | *Yes* |
| *Observations* | 5,777 | 5,777 | 5,777 | 5,777 |
| *Number of firms* | 1,483 | 1,483 | 1,483 | 1,483 |
| $R^2$ | 0.049 | 0.085 | 0.053 | 0.089 |

This table reports the estimated results from the regressions of robust test of the quantity of green innovation on stock price crash risk. The dependent variable, NCKEW and DUVOL, is measured as of year t+1, while the independent variables are measured as of year t. For the fixed effect model, t statistic based on the robust standard error is in parentheses, and for the GMM model, z statistic based on the robust standard error is in parentheses. ***, **, and * indicate significance at the 1%, 5%, and 10% levels, respectively.

Table 5-2 Robustness tests of the quantity of green innovation on stock price crash risk

| *Model* | *GMM* | |
|---|---|---|
| | (5) | (6) |
| *VARIABLES* | $NCSKEW_{i,t+1}$ | $DUVOL_{i,t+1}$ |
| $NCSKEW_{i,t}$ | 0.0418 | |
| | (1.6259) | |
| $DUVOL_{i,t+1}$ | | 0.0527* |
| | | (1.6602) |
| $NGInnovation2_{i,t}$ | 0.0215* | 0.0176* |
| | (1.8292) | (1.6494) |
| $Oturnover_{i,t}$ | -0.0583 | -0.0111 |
| | (-1.2133) | (-0.3530) |
| $Sigma_{i,t}$ | 5.2811*** | 2.0262** |
| | (3.6027) | (2.2978) |
| $Ret_{i,t}$ | 3.5974 | 3.6715** |
| | (1.4986) | (2.2209) |
| $Size_{i,t}$ | 0.2143** | 0.1440** |
| | (2.4488) | (2.3519) |
| $BM_{i,t}$ | -0.9157*** | -0.7117*** |
| | (-4.2699) | (-5.2860) |
| $Lev_{i,t}$ | -0.4518* | -0.3366* |
| | (-1.6475) | (-1.7307) |
| $Roa_{i,t}$ | -0.3533 | -0.4019** |
| | (-1.2708) | (-2.0860) |
| $ABACC_{i,t}$ | 0.5415* | 0.4894** |
| | (1.7101) | (2.2538) |
| $Tophold_{i,t}$ | 0.3399 | 0.2271 |
| | (0.6263) | (0.5836) |
| $Power_{i,t}$ | -0.0067 | -0.0659 |
| | (-0.0860) | (-1.4047) |
| $Maghold_{i,t}$ | -0.0207 | -0.0307 |
| | (-0.2870) | (-0.6684) |
| $Balance_{i,t}$ | 0.1707** | 0.1121** |
| | (2.3362) | (2.4120) |
| $Inshold_{i,t}$ | -0.0744 | -0.0155 |
| | (-1.4726) | (-0.4324) |
| *Firm FE* | *Yes* | *Yes* |
| *Year FE* | *Yes* | *Yes* |
| *Observations* | 3,658 | 3,658 |
| *Number of firms* | 1,044 | 1,044 |
| *AR*(1) | -11.38*** (Pr > z = 0.000) | -12.01 (Pr > z = 0.000) |
| *AR*(2) | 0.91 (Pr > z = 0.360) | 0.90 (Pr > z = 0.367) |
| *Sargan test* | 106.41 (Prob > chi2 = 0.525) | 99.01 (Prob > chi2 = 0.720) |

This table reports the estimated results from the regressions of robust test of the quantity of green innovation on stock price crash risk. The dependent variable, NCKEW and DUVOL, is measured as of year t+1, while the independent variables are measured as of year t. For the fixed effect model, t statistic based on the robust standard error is in parentheses, and for the GMM model, z statistic based on the robust standard error is in parentheses. ***, **, and * indicate significance at the 1%, 5%, and 10% levels, respectively.

Table 5-3 Robustness tests of the quantity of green innovation on stock price crash risk

| *Model* | *FE* | | | *GMM* |
|---|---|---|---|---|
| | (1) | (2) | | (3) |
| *VARIABLES* | $DUVOL_{i,t+1}$ | $DUVOL_{i,t+1}$ | *VARIABLES* | $DUVOL_{i,t+1}$ |
| $DUVOL_{i,t}$ | | -0.1229*** | $DUVOL_{i,t}$ | 0.0524* |
| | | (-8.2299) | | (1.6461) |
| $NGInnovation1_{i,t}$ | 0.0003* | 0.0002 | $NGInnovation1_{i,t}$ | 0.0006 |
| | (1.6808) | (1.0968) | | (1.6144) |
| $Oturnover_{i,t}$ | | -0.0151 | $Oturnover_{i,t}$ | -0.0111 |
| | | (-0.6746) | | (-0.3533) |
| $Sigma_{i,t}$ | | -0.7256 | $Sigma_{i,t}$ | 2.0639** |
| | | (-1.2653) | | (2.3455) |
| $Ret_{i,t}$ | | 3.1316*** | $Ret_{i,t}$ | 3.6177** |
| | | (2.7388) | | (2.1928) |
| $Size_{i,t}$ | | 0.0646** | $Size_{i,t}$ | 0.1454** |
| | | (2.4418) | | (2.3700) |
| $BM_{i,t}$ | | -0.3825*** | $BM_{i,t}$ | -0.7098*** |
| | | (-4.8186) | | (-5.2660) |
| $Lev_{i,t}$ | | -0.2916*** | $Lev_{i,t}$ | -0.3423* |
| | | (-2.9578) | | (-1.7667) |
| $Roa_{i,t}$ | | -0.3862** | $Roa_{i,t}$ | -0.4026** |
| | | (-2.5747) | | (-2.0943) |
| $ABACC_{i,t}$ | | 0.4666*** | $ABACC_{i,t}$ | 0.4969** |
| | | (3.1757) | | (2.2728) |
| $Tophold_{i,t}$ | | 0.0287 | $Tophold_{i,t}$ | 0.2273 |
| | | (0.1636) | | (0.5843) |
| $Power_{i,t}$ | | 0.0159 | $Power_{i,t}$ | -0.0677 |
| | | (0.5561) | | (-1.4437) |
| $Maghold_{i,t}$ | | -0.0242 | $Maghold_{i,t}$ | -0.0314 |
| | | (-0.8543) | | (-0.6821) |
| $Balance_{i,t}$ | | 0.0711** | $Balance_{i,t}$ | 0.1137** |
| | | (2.3732) | | (2.4492) |
| $Inshold_{i,t}$ | | 0.0250 | $Inshold_{i,t}$ | -0.0151 |
| | | (1.0831) | | (-0.4217) |
| $INDUSTRYID_{i,t}$ | | -1.2991** | $INDUSTRYID_{i,t}$ | 0.0006 |
| | | (-2.3384) | | (1.6144) |
| *Constant* | -0.3534*** | -0.1229*** | | |
| | (-8.3320) | (-8.2299) | | |
| *Firm FE* | *Yes* | *Yes* | *Firm FE* | *Yes* |
| *Year FE* | *Yes* | *Yes* | *Year FE* | *Yes* |
| | | | *Observations* | 3,658 |
| | | | *Number of firms* | 1,044 |
| *Observations* | 5,777 | 5,777 | *AR*(1) | -11.96 (Pr > z = 0.000) |
| *Number of firms* | 1,483 | 1,483 | *AR*(2) | 0.88 (Pr > z = 0.380) |
| $R^2$ | 0.054 | 0.089 | *Sargan test* | 100.32 (Prob > chi2 = 0.688) |

This table reports the robustness tests of the estimated results from the regressions of the quantity of green innovation on stock price crash risk. The dependent variable, DUVOL, is measured as of year t+1, while the independent variables are measured as of year t. For the fixed effect model, t statistic based on the robust standard error is in parentheses, and for the GMM model, z statistic based on the robust standard error is in parentheses. ***, **, and * indicate significance at the 1%, 5%, and 10% levels, respectively.

Table 6 Robustness tests of the effect of the quality of green innovation on stock price crash risk

| *Model* | *FE* | | | *GMM* |
|---|---|---|---|---|
| | (1) | (2) | | (3) |
| *VARIABLES* | $DUVOL_{i,t+1}$ | $DUVOL_{i,t+1}$ | *VARIABLES* | $DUVOL_{i,t+1}$ |
| $DUVOL_{i,t}$ | | -0.1226*** | $DUVOL_{i,t}$ | 0.0578* |
| | | (-8.2310) | | (1.8295) |
| $QGInnovation_{i,t}$ | -0.0227*** | -0.0249*** | $QGInnovation_{i,t}$ | -0.0302*** |
| | (-3.5177) | (-3.9046) | | (-4.0145) |
| $Oturnover_{i,t}$ | | -0.0160 | $Oturnover_{i,t}$ | -0.0193 |
| | | (-0.7158) | | (-0.6309) |
| $Sigma_{i,t}$ | | -0.7865 | $Sigma_{i,t}$ | 2.2159** |
| | | (-1.3775) | | (2.5629) |
| $Ret_{i,t}$ | | 3.1779*** | $Ret_{i,t}$ | 3.5872** |
| | | (2.7843) | | (2.1751) |
| $Size_{i,t}$ | | 0.0683*** | $Size_{i,t}$ | 0.0933 |
| | | (2.6044) | | (1.6064) |
| $BM_{i,t}$ | | -0.3882*** | $BM_{i,t}$ | -0.6492*** |
| | | (-4.8976) | | (-4.8599) |
| $Lev_{i,t}$ | | -0.3012*** | $Lev_{i,t}$ | -0.2943 |
| | | (-3.0660) | | (-1.5867) |
| $Roa_{i,t}$ | | -0.3942*** | $Roa_{i,t}$ | -0.3552* |
| | | (-2.6615) | | (-1.8803) |
| $ABACC_{i,t}$ | | 0.4784*** | $ABACC_{i,t}$ | 0.5411** |
| | | (3.2720) | | (2.4326) |
| $Tophold_{i,t}$ | | 0.0476 | $Tophold_{i,t}$ | 0.2342 |
| | | (0.2724) | | (0.5974) |
| $Power_{i,t}$ | | 0.0157 | $Power_{i,t}$ | -0.0918* |
| | | (0.5497) | | (-1.7897) |
| $Maghold_{i,t}$ | | -0.0285 | $Maghold_{i,t}$ | -0.0239 |
| | | (-1.0061) | | (-0.5035) |
| $Balance_{i,t}$ | | 0.0753** | $Balance_{i,t}$ | 0.1196** |
| | | (2.5208) | | (2.5054) |
| $Inshold_{i,t}$ | | 0.0245 | $Inshold_{i,t}$ | -0.0367 |
| | | (1.0601) | | (-0.9969) |
| *Constant* | -0.3530*** | -1.3683** | | |
| | (-8.2831) | (-2.4867) | | |
| *Firm FE* | *Yes* | *Yes* | *Firm FE* | *Yes* |
| *Year FE* | *Yes* | *Yes* | *Year FE* | *Yes* |
| | | | *Observations* | 3,658 |
| | | | *Number of firms* | 1,044 |
| *Observations* | 5,777 | 5,777 | *AR*(1) | -12.25 (Pr > z = 0.000) |
| *Number of firms* | 1,483 | 1,483 | *AR*(2) | 0.91 (Pr > z = 0.360) |
| $R^2$ | 0.055 | 0.091 | *Sargan test* | 117.00 (Prob > chi2 = 0.261) |

This table reports the robustness tests of the estimated results from the regressions of the quality of green innovation on stock price crash risk. The dependent variable, DUVOL, is measured as of year t+1, while the independent variables are measured as of year t. For the fixed effect model, t statistic based on the robust standard error is in parentheses, and for the GMM model, z statistic based on the robust standard error is in parentheses. ***, **, and * indicate significance at the 1%, 5%, and 10% levels, respectively.

## 5| ADDRESSING ENDOGENEITY CONCERNS

Our analysis proves that there is a negative correlation between the quality of green innovation and the risk of stock crashes. However, when the relationship between green innovation and stock crash risk is affected by unobserved company-level characteristics, an endogeneity problem that affects the effectiveness of the model estimation results will appear. To solve the endogeneity problem to a certain extent, we use external policy shocks to perform a DID regression.

With the continuous improvement of the industrialization process since China's reform and opening up, carbon emissions have been maintained at a high level, which has become the main cause of air pollution in China. To control carbon emissions, the introduction and implementation of relevant laws and regulations have attracted much attention. In June 2013, China creatively introduced the carbon emission trading mechanism originated from the developed capital markets in Europe and America, and successively established carbon exchanges in Shenzhen, Beijing, Shanghai and other provinces and five cities to carry out pilot trading. After four years of pilot trading experiments, a unified market for carbon emissions trading was officially launched in China at the end of 2017. As of March 2021, the carbon market in the carbon trading pilot area covers more than 20 industries such as steel, electric power and cement, nearly 3000 key emission enterprises, covering 440 million tons of carbon emissions, with a cumulative turnover of about 10.47 billion yuan. Within the pilot scope, the total amount and intensity of carbon emissions of enterprises have achieved "double reduction", which shows the effect of carbon market controlling carbon emissions at a lower cost.

Relevant studies show that the emission trading pilot promotes green technology innovation activities of regional enterprises, and Zhang et al. (2019) show that the emission trading pilot significantly improves the level of green technology innovation of the pilot area through quasi-natural experimental research. Table 7-1 shows that the impact of carbon emission trading right policy on the quality of green technology innovation is positive and significant at the 5% confidence level, indicating the rationality of the selection of instrumental variables. In this way, the exogenous policy impact of carbon emission trading rights can be used as a instrumental variable of green technology innovation, which can effectively alleviate the endogeneity problems involved in this study and the deviation caused.

Considering that there is a lag period of the explained variable in the research model, the $NCSKEW_{i,t}(DUVOL_{i,t})$ is endogenous, So the GMM method is used to estimate the model parameters by using the multiple lag period of the $NCSKEW_{i,t}(DUVOL_{i,t})$ and exogenous policy impact as the instrumental variables of endogenous control variable $NCSKEW_{i,t}(DUVOL_{i,t})$ and independent variable $QGInnovation_{i,t}$ .(The endogenous variables in the model are $NCSKEW_{i,t}(DUVOL_{i,t})$ and $QGInnovation_{i,t}$ ,the instrumental variable is the lag period of the $NCSKEW_{i,t}(DUVOL_{i,t})$ and the exogenous policy impact)

The model of the difference-in-differences regression is specified as follows (4):

(4)

$$NCSKEW_{i,t+1}\left(DUVOL_{i,t+1}\right)=\beta_0+\beta_1 QGInnovation_{i,t}+\beta_2 Policy_i Post_{i,t}+\sum\beta_j\left(Control\ variables\right)_{i,t}+\sum Firm+\sum Year+\varepsilon_{i,t}$$

"Post" is a dummy variable, since the establishment of seven carbon markets is concentrated in the second half of 2013 and the first half of 2014, taking 2014 as the benchmark, if it is in and after 2014, the value of "Post" is 1, otherwise, the value of "Post" is 0. "Policy" is a dummy variable whose value equals one if the company is located in the policy pilot area and zero otherwise. If green innovation can alleviate stock crash risk, we expect this coefficient β 1 to continue to be negative and significant.

According to the regression results in Table 7-2(1), the estimated results of the regression are in the coefficient table β 1 and were negative and significant at the 5% confidence level. Moreover, in the robustness test results in Table 7-2(2), the correlation coefficient is consistent with the benchmark regression, and β1 is significant at the 1% confidence level. The regression results show that after trying to eliminate the impact of endogeneity problems on the model results, the impact of green technology innovation quality and stock price crash risk is still negative. In addition, the regression results of the control variables are basically consistent with expectations.

Table 7-1 Difference-in-differences (DID) regressions.

| | *DID* | | *DID* |
|---|---|---|---|
| | (1) | | (2) |
| *VARIABLES* | $QGInnovation_{i,t}$ | *VARIABLES* | $QGInnovation_{i,t}$ |
| $NCSKEW_{i,t}$ | -0.0003 | $DUVOL_{i,t}$ | 0.0162 |
| | (-0.0137) | | (0.5407) |
| $policypost_{i,t}$ | 0.1474** | $policypost_{i,t}$ | 0.1477** |
| | (2.3773) | | (2.3820) |
| $Oturnover_{i,t}$ | -0.0480 | $Oturnover_{i,t}$ | -0.0482 |
| | (-1.1283) | | (-1.1345) |
| $Sigma_{i,t}$ | -1.1065 | $Sigma_{i,t}$ | -1.0466 |
| | (-1.0094) | | (-0.9564) |
| $Ret_{i,t}$ | 0.5354 | $Ret_{i,t}$ | 0.7158 |
| | (0.2268) | | (0.3025) |
| $Size_{i,t}$ | 0.0128 | $Size_{i,t}$ | 0.0130 |
| | (0.2536) | | (0.2568) |
| $BM_{i,t}$ | -0.1214 | $BM_{i,t}$ | -0.1173 |
| | (-0.8225) | | (-0.7953) |
| $Lev_{i,t}$ | -0.1628 | $Lev_{i,t}$ | -0.1635 |
| | (-0.9047) | | (-0.9087) |
| $Roa_{i,t}$ | -0.2017 | $Roa_{i,t}$ | -0.2016 |
| | (-0.7236) | | (-0.7231) |
| $ABACC_{i,t}$ | 0.5272* | $ABACC_{i,t}$ | 0.5287* |
| | (1.8309) | | (1.8360) |
| $Tophold_{i,t}$ | 0.4836 | $Tophold_{i,t}$ | 0.4853 |
| | (1.4840) | | (1.4894) |
| $Power_{i,t}$ | -0.0468 | $Power_{i,t}$ | -0.0474 |
| | (-0.7943) | | (-0.8047) |
| $Maghold_{i,t}$ | -0.1227** | $Maghold_{i,t}$ | -0.1229** |
| | (-2.2056) | | (-2.2086) |
| $Balance_{i,t}$ | 0.1590*** | $Balance_{i,t}$ | 0.1586*** |
| | (2.7947) | | (2.7875) |
| $Inshold_{i,t}$ | -0.0190 | $Inshold_{i,t}$ | -0.0197 |
| | (-0.4071) | | (-0.4225) |
| *Firm FE* | *Yes* | *Firm FE* | *Yes* |
| *Year FE* | *Yes* | *Year FE* | *Yes* |

The table reports the estimated results from the difference-in-difference regressions of policy on the quality of green innovation in matching samples from 2008 to 2018.t statistic based on the robust standard error is in parentheses. ***, **, and * indicate significance at the 1%, 5%, and 10% levels, respectively.

Table 7-2 Difference-in-differences (DID) regressions.

| | *DID* | | *DID* |
|---|---|---|---|
| | (1) | | (2) |
| *VARIABLES* | $NCSKEW_{i,t+1}$ | *VARIABLES* | $DUVOL_{i,t+1}$ |
| $NCSKEW_{i,t}$ | 0.0406 | $DUVOL_{i,t}$ | 0.0560* |
| | (1.6307) | | (1.7741) |
| $QGInnovation_{i,t}$ | -0.0303** | $QGInnovation_{i,t}$ | -0.0302*** |
| | (-2.5268) | | (-3.8457) |
| $policypost_{i,t}$ | -0.1895** | $policypost_{i,t}$ | -0.1550** |
| | (-1.9792) | | (-2.2403) |
| $Oturnover_{i,t}$ | -0.0474 | $Oturnover_{i,t}$ | -0.0202 |
| | (-1.0055) | | (-0.6642) |
| $Sigma_{i,t}$ | 4.0734*** | $Sigma_{i,t}$ | 2.2415*** |
| | (2.9611) | | (2.6032) |
| $Ret_{i,t}$ | 6.0901** | $Ret_{i,t}$ | 3.4848** |
| | (2.5394) | | (2.1313) |
| $Size_{i,t}$ | 0.1750** | $Size_{i,t}$ | 0.0954* |
| | (2.0636) | | (1.6506) |
| $BM_{i,t}$ | -0.7797*** | $BM_{i,t}$ | -0.6498*** |
| | (-3.7065) | | (-4.8791) |
| $Lev_{i,t}$ | -0.5409** | $Lev_{i,t}$ | -0.2859 |
| | (-2.0379) | | (-1.5405) |
| $Roa_{i,t}$ | -0.4550* | $Roa_{i,t}$ | -0.3389* |
| | (-1.6737) | | (-1.7846) |
| $ABACC_{i,t}$ | $0.5315^{*}$ | $ABACC_{i,t}$ | $0.5397^{**}$ |
| | (1.6981) | | (2.4409) |
| $Tophold_{i,t}$ | 0.3471 | $Tophold_{i,t}$ | 0.2131 |
| | (0.6383) | | (0.5441) |
| $Power_{i,t}$ | -0.0023 | $Power_{i,t}$ | $-0.0860^{*}$ |
| | (-0.0289) | | (-1.6806) |
| $Maghold_{i,t}$ | 0.0178 | $Maghold_{i,t}$ | -0.0279 |
| | (0.2407) | | (-0.5869) |
| $Balance_{i,t}$ | $0.1577^{**}$ | $Balance_{i,t}$ | $0.1238^{***}$ |
| | (2.2822) | | (2.5903) |
| $Inshold_{i,t}$ | -0.0764 | $Inshold_{i,t}$ | -0.0376 |
| | (-1.4651) | | (-1.0248) |
| *Firm FE* | *Yes* | *Firm FE* | *Yes* |
| *Year FE* | *Yes* | *Year FE* | *Yes* |
| *Observations* | 3,658 | *Observations* | 3,658 |
| *Number of firms* | 1,044 | *Number of firms* | 1,044 |
| *AR*(1) | -11.44*** | *AR*(1) | -12.23*** |
| | (Pr > z = 0.000) | | (Pr > z = 0.000) |
| *AR*(2) | 1.01 | *AR*(2) | 0.94 |
| | (Pr > z = 0.312) | | (Pr > z = 0.345) |
| *Sargan test* | 107.82 | *Sargan test* | 116.94 |
| | (Prob > chi2 = 0.487) | | (Prob > chi2 = 0.262) |

The table reports the estimated results from the difference-in-difference regressions of stock crash risk in matching samples from 2008 to 2018.z statistic based on the robust standard error is in parentheses. ***, **, and * indicate significance at the 1%, 5%, and 10% levels, respectively.

## 6| THE IMPACT OF THE NATURE OF ENTERPRISE OWNERSHIP

The samples are grouped based on whether they are state-owned enterprises or not. According to the regression results in Tables 8 (1) and (2), in the nonstate-owned enterprise sample, the quality of green technology innovation is negatively correlated with the risk of a stock price collapse and is significant at the 1% confidence level. In the sample of state-owned enterprises, the relationship between the quality of green technological innovation and the risk of a stock price collapse is positively correlated and is significant at the 10% confidence level, it is very weak and not significant in the robustness test in Table 9 (2). We believe that state-owned enterprises usually have a certain relationship with the government, and it is easier to obtain various resources. Regional state-owned enterprises often shoulder part of the development goals of regional governments. Due to the close relationship between state-owned enterprises and the government, the development of state-owned enterprises cannot be separated from the support of the government, which makes it possible for state-owned enterprises to engage in malpractice for personal gain and opens up one side of the law in terms of relevant supervision and law enforcement. State-owned Enterprises can, thus, avoid environmental regulation and other pollution fines and confiscations. In this way, the behavior of improving the quality of green technology innovation adopted by state-owned enterprises will not be regarded as management self-discipline, but instead will be viewed as performance related to declining political resources; that is, state-owned enterprises gradually lose the government's resource tilt and nepotism in the case of government assistance recession. The reversal of political achievements and strengthening of information-hiding behavior may cause a difference between information disclosure and the real time window, increasing the risk of a stock price crash. In general, the special nature of state-owned enterprises distorts the original negative relationship between the quality of green technology innovation and the stock price collapse. In general, the relationship between the two is not significant, and there is a positive correlation to a certain extent. Thus, Hypothesis 3 holds. According to the robustness test results in columns (1) and (2) of Table 9, the relevant results are consistent with the benchmark regression results in Table 8, indicating the reliability of the regression results. In addition, the regression results of the control variables are basically consistent with expectations.

Table 8 Quality of green innovation and stock crash risk in SOEs and non-SOEs

| *MODEL* | *GMM* | |
|---|---|---|
| | (1) | (2) |
| | *Non*-SOEs | SOEs |
| *VARIABLES* | $NCSKEW_{i,t+1}$ | $NCSKEW_{i,t+1}$ |
| $NCSKEW_{i,t}$ | 0.0318 | -0.0137 |
| | (1.0112) | (-0.4111) |
| $QGInnovation_{i,t}$ | -0.0336*** | 0.1054* |
| | (-5.6149) | (1.9511) |
| $Oturnover_{i,t}$ | -0.0649 | 0.0156 |
| | (-1.2623) | (0.1673) |
| $Sigma_{i,t}$ | 3.4537* | 4.1960** |
| | (1.8563) | (1.9620) |
| $Ret_{i,t}$ | 5.3942* | 2.8834 |
| | (1.6972) | (0.7105) |
| $Size_{i,t}$ | 0.3302*** | 0.0288 |
| | (3.0254) | (0.1819) |
| $BM_{i,t}$ | -0.8598** | -0.8714*** |
| | (-2.5459) | (-3.1865) |
| $Lev_{i,t}$ | -0.6889** | 0.0504 |
| | (-1.9832) | (0.1165) |
| $Roa_{i,t}$ | -0.8478*** | -0.0537 |
| | (-2.6103) | (-0.0746) |
| $ABACC_{i,t}$ | 0.3389 | -0.0335 |
| | (0.9317) | (-0.0639) |
| $Tophold_{i,t}$ | 1.1378 | -0.9120 |
| | (1.5387) | (-1.1735) |
| $Power_{i,t}$ | -0.0902 | 0.0254 |
| | (-0.8566) | (0.2097) |
| $Maghold_{i,t}$ | -0.1014 | 0.1133 |
| | (-1.0920) | (0.9576) |
| $Balance_{i,t}$ | 0.2325*** | 0.0304 |
| | (2.7465) | (0.2587) |
| $Inshold_{i,t}$ | -0.0900 | -0.0724 |
| | (-1.2214) | (-0.8985) |
| *Firm FE* | *Yes* | *Yes* |
| *Year FE* | *Yes* | *Yes* |
| *Observations* | 2,022 | 1,607 |
| *Number of firms* | 654 | 404 |
| *AR*(1) | -7.62 | -7.93 |
| | (Pr > z = 0.000) | (Pr > z = 0.000) |
| *AR*(2) | 0.40 | 0.00 |
| | (Pr > z = 0.693) | (Pr > z = 0.999) |
| *Sargan test* | 84.45 | 124.10 |
| | (Prob > chi2 = 0.955) | (Prob > chi2 = 0.138) |

This table reports the estimated results from the regressions of stock crash risk on the quality of green innovation in SOE and non-SOE samples. The dependent variable, NCSKEW, is measured as of year t+1, while the independent variables are measured as of year t. z statistic based on the robust standard error is in parentheses. ***, ** and * indicate significance at the 1%, 5%, and 10% levels, respectively.

Table 9 Robustness tests of the quality of green innovation and stock crash risk in SOEs and non-SOEs

| *MODEL* | *GMM* | |
|---|---|---|
| | （1） | （2） |
| | *Non*-SOEs | SOEs |
| *VARIABLES* | $DUVOL_{i,t+1}$ | $DUVOL_{i,t+1}$ |
| $DUVOL_{i,t}$ | 0.0373 | 0.0032 |
| | (1.1028) | (0.0830) |
| $QGInnovation_{i,t}$ | -0.0284** | 0.0580 |
| | (-2.2434) | (1.5832) |
| $Oturnover_{i,t}$ | 0.0009 | -0.0121 |
| | (0.0247) | (-0.1981) |
| $Sigma_{i,t}$ | 0.9066 | 0.1179 |
| | (0.7277) | (0.0774) |
| $Ret_{i,t}$ | 4.0762** | 6.0743** |
| | (1.9694) | (2.0332) |
| $Size_{i,t}$ | 0.1168 | 0.1588 |
| | (1.5571) | (1.2757) |
| $BM_{i,t}$ | -0.5399*** | -0.7383*** |
| | (-2.7887) | (-3.6505) |
| $Lev_{i,t}$ | -0.5905** | 0.0483 |
| | (-2.5219) | (0.1601) |
| $Roa_{i,t}$ | -0.5850*** | -0.3977 |
| | (-2.6913) | (-0.9230) |
| $ABACC_{i,t}$ | 0.4238 | 0.4375 |
| | (1.4995) | (1.2478) |
| $Tophold_{i,t}$ | 0.9922* | -1.1864** |
| | (1.9131) | (-2.1672) |
| $Power_{i,t}$ | -0.0970 | -0.0804 |
| | (-1.4612) | (-0.9953) |
| *$Maghold_{i,t}$* | -0.0803 | 0.0661 |
| | (-1.2429) | (0.8149) |
| *$Balance_{i,t}$* | 0.1418** | -0.0088 |
| | (2.2794) | (-0.1135) |
| *$Inshold_{i,t}$* | -0.0157 | -0.0447 |
| | (-0.2978) | (-0.7685) |
| | | |
| *Firm FE* | *Yes* | *Yes* |
| *Year FE* | *Yes* | *Yes* |
| *Observations* | 2,022 | 1,607 |
| *Number of firms* | 654 | 404 |
| *AR*(1) | -9.13*** | -8.32*** |
| | (Pr > z = 0.000) | (Pr > z = 0.000) |
| *AR*(2) | -0.36 | 0.91 |
| | (Pr > z = 0.716) | (Pr > z = 0.362) |
| *Sargan test* | 87.83 | 122.00 |
| | (Prob > chi2 = 0.923) | (Prob > chi2 = 0.169) |

This table reports the robustness tests of the estimated results from the regressions of stock crash risk on the quality of green innovation in SOE and non-SOE samples. The dependent variable, DUVOL, is measured as of year t+1, while the independent variables are measured as of year t. z statistic based on the robust standard error is in parentheses. ***, ** and * indicate significance at the 1%, 5%, and 10% levels, respectively.

## 7| MECHANISM OF THE IMPACT OF QUALITY OF GREEN TECHNOLOGY INNOVATION

Research and analysis show that the quality of green technology innovation reduces the risk of a stock price collapse. In this section, we explore the internal mechanism of the relationship between the quality of green technological innovation and stock price crashes. As the share price collapse is mainly due to the concealment of negative information by the management(Jin & Myers, 2006; Hutton et al., 2009)，This section focuses on the number of negative media reports of enterprises. Our theory assumes that the quality of green technology innovation will enhance the accuracy and timeliness of company information disclosure through both information and shareholderism. This will further reduce the concealment of negative information by the management, so that the overall number of negative news reports of the company will be relatively higher. Based on this, we believe that the improvement of the quality of green technology innovation will increase the exposure and disclosure of negative information about the company.

To test this hypothesis, we collected relevant data from the financial news reports of listed companies from the CNRDS database. The sources of online financial news include news report data from more than 400 important online media outlets. Among them, the most noteworthy are news reports of 20 important online financial media: Hexun.com, Sina Finance, Eastern Fortune.Net, Tencent Finance, NetEase Finance, Phoenix Finance, China Economic Net, Sohu Finance, Financial Circle, EC Finance, FT Chinese Net, Panorama Net, China Finance Online, China Securities Net, Securities Star, Caixin Net, The Paper Net, China Business News, 21CN Financial Channel, and Caijing.com. The abovementioned online financial media outlets rank at the forefront of China in terms of the quantity and quality of reports, and they are also the online media that investors and stakeholders consult most frequently and have high reference significance and value.

Since the stock price collapse risk involved in this research is closely related to whether the company’s management has hidden relevant bad news, our research focuses on the impact of the quality of green technology innovation on the amount of original negative news of the company and the amount of all negative news to explore whether media reports play an intermediary role in the relationship between the quality of green technology innovation and stock price crash. We establish a fixed effects model to analyze the intermediary mechanism. Model (5) discusses the relationship between the quality of green technology innovation and the number of negative company news items, and model (6) discusses the relationship between the number of negative company news items and stock price crashes.

（5）

$$ONMEDIA_{i,t}(ANMEDIA_{i,t}) = \beta_0 + \beta_1 QGInnovation_{i,t} + \sum \beta_j \left(Control\ variables\right)_{i,t} + \sum Firm + \sum Year + \varepsilon_{i,t}$$

（6）

$$NCSKEW_{i,t+1} = \beta_0 + \beta_1 ONMEDIA_{i,t}(ANMEDIA_{i,t}) + \sum \beta_j \left(Control\ variables\right)_{i,t} + \sum Firm + \sum Year + \varepsilon_{i,t}$$

According to the regression results in columns (1) and (2) of Table 10, the green technology innovation quality index is positively correlated with the number of original negative news reports by listed companies and is significant at the 5% confidence level; the number of original negative news reports by listed companies is negatively correlated with the risk of a stock price collapse and significant at the 1% confidence level, and H4a and H5a are valid. The quality of corporate green technology innovation behavior indicates the management principle of self-sacrifice and respect for the rights and interests of shareholders, which increases the exposure of a company's information and ultimately enables management to make public reports about the company's own negative conditions through the media according to the real time window, effectively communicating information to stakeholders. The negative news reports of listed companies also play a role in reducing the degree of information asymmetry with stakeholders and ultimately curb the risk of a stock price collapse. To ensure the robustness of the research results, we use the total number of negative news reports of listed companies to replace the number of original negative news reports of listed companies.

According to the robustness test results in columns (1) and (2) of Table 11, it is consistent with the basic regression results, indicating that the core regression results of this study are robust. In addition, the regression results of the control variables are basically consistent with expectations.

Table 10 Quality of green innovation on stock crash risk through negative media reports

| *Model* | *FE* | | *FE* |
|---|---|---|---|
| | (1) | | (2) |
| *VARIABLES* | $ONMEDIA_{i,t}$ | VARIABLES | $NCSKEW_{i,t+1}$ |
| $QGInnovation_{i,t}$ | 1.1993** | $NCSKEW_{i,t}$ | -0.1271*** |
| | (1.9970) | | (-8.5695) |
| | | $ONMEDIA_{i,t}$ | -0.0012*** |
| | | | (-3.4282) |
| $Oturnover_{i,t}$ | 1.7342 | $Oturnover_{i,t}$ | -0.0303 |
| | (1.2493) | | (-0.8629) |
| $Sigma_{i,t}$ | 314.6518*** | $Sigma_{i,t}$ | -0.4621 |
| | (3.9591) | | (-0.5383) |
| $Ret_{i,t}$ | -346.0334*** | $Ret_{i,t}$ | 4.0806** |
| | (-3.5603) | | (2.4414) |
| $Size_{i,t}$ | 11.8545*** | $Size_{i,t}$ | 0.1347*** |
| | (6.0308) | | (3.4243) |
| $BM_{i,t}$ | -22.6986*** | $BM_{i,t}$ | -0.6292*** |
| | (-5.1022) | | (-5.1004) |
| $Lev_{i,t}$ | -7.5913 | $Lev_{i,t}$ | -0.5172*** |
| | (-1.3695) | | (-3.5501) |
| $Roa_{i,t}$ | -12.3375 | $Roa_{i,t}$ | -0.6147*** |
| | (-1.3548) | | (-2.8362) |
| $ABACC_{i,t}$ | 12.1986 | $ABACC_{i,t}$ | 0.4090* |
| | (1.5286) | | (1.8754) |
| $Tophold_{i,t}$ | -1.1697 | $Tophold_{i,t}$ | 0.1568 |
| | (-0.1353) | | (0.6041) |
| $Power_{i,t}$ | -0.5171 | $Power_{i,t}$ | 0.0364 |
| | (-0.3715) | | (0.8073) |
| $Maghold_{i,t}$ | -3.2370 | $Maghold_{i,t}$ | -0.0245 |
| | (-1.0615) | | (-0.5956) |
| $Balance_{i,t}$ | -2.4325 | $Balance_{i,t}$ | 0.1188*** |
| | (-0.9684) | | (2.6285) |
| $Inshold_{i,t}$ | -2.4016 | $Inshold_{i,t}$ | 0.0283 |
| | (-1.1522) | | (0.8393) |
| *Firm FE* | Yes | *Firm FE* | Yes |
| *Year FE* | Yes | *Year FE* | Yes |
| *Constant* | -255.8873*** | *Constant* | -2.7564*** |
| | (-5.9275) | | (-3.364) |
| *Observations* | 5,775 | *Observations* | 5,775 |
| *Number of firms* | 1,483 | *Number of firms* | 1,483 |
| $R^2$ | 0.070 | $R^2$ | 0.087 |

This table reports the estimated results from the regressions of the quality of green innovation on the number of original negative news reports by listed companies and the number of original negative news reports by listed companies on stock crash risk. t statistic based on the robust standard error is in parentheses. ***, ** and * indicate significance at the 1%, 5%, and 10% levels, respectively.

Table 11 Robustness tests of the effect of the quality of green innovation on stock crash risk through negative media reports

| *Model* | FE | | FE |
|---|---|---|---|
| | (1) | | (2) |
| *VARIABLES* | $ANMEDIA_{i,t}$ | *VARIABLES* | $NCSKEW_{i,t+1}$ |
| $QGInnovation_{i,t}$ | 5.4822** | $NCSKEW_{i,t}$ | -0.1267*** |
| | (2.0297) | | (-8.5333) |
| | | $ANMEDIA_{i,t}$ | -0.0002*** |
| | | | (-3.2602) |
| $Oturnover_{i,t}$ | 4.5017 | $Oturnover_{i,t}$ | -0.0315 |
| | (0.6352) | | (-0.8965) |
| $Sigma_{i,t}$ | 1664.3236*** | $Sigma_{i,t}$ | -0.5191 |
| | (4.1300) | | (-0.6060) |
| $Ret_{i,t}$ | -1608.9489*** | $Ret_{i,t}$ | 4.1937** |
| | (-2.8484) | | (2.5223) |
| $Size_{i,t}$ | 41.3558*** | $Size_{i,t}$ | 0.1284*** |
| | (3.7518) | | (3.2460) |
| $BM_{i,t}$ | -75.6650*** | $BM_{i,t}$ | -0.6163*** |
| | (-2.8868) | | (-5.0137) |
| $Lev_{i,t}$ | -25.0878 | $Lev_{i,t}$ | -0.5128*** |
| | (-0.7575) | | (-3.5123) |
| $Roa_{i,t}$ | -76.0541* | $Roa_{i,t}$ | -0.6142*** |
| | (-1.7228) | | (-2.8558) |
| $ABACC_{i,t}$ | 94.6072** | $ABACC_{i,t}$ | 0.4121* |
| | (2.0601) | | (1.8910) |
| $Tophold_{i,t}$ | 8.3415 | $Tophold_{i,t}$ | 0.1596 |
| | (0.1762) | | (0.6175) |
| $Power_{i,t}$ | 3.5673 | $Power_{i,t}$ | 0.0377 |
| | (0.3575) | | (0.8341) |
| $Maghold_{i,t}$ | -5.4747 | $Maghold_{i,t}$ | -0.0217 |
| | (-0.3843) | | (-0.5249) |
| $Balance_{i,t}$ | 8.2184 | $Balance_{i,t}$ | 0.1231*** |
| | (0.5329) | | (2.7311) |
| $Inshold_{i,t}$ | -19.7756 | $Inshold_{i,t}$ | 0.0274 |
| | (-1.3076) | | (0.8118) |
| *Firm FE* | *Yes* | *Firm FE* | *Yes* |
| *Year FE* | *Yes* | *Year FE* | *Yes* |
| *Constant* | -1022.3349*** | *Constant* | -2.6587*** |
| | (-4.2382) | | (-3.2551) |
| *Observations* | 5,775 | *Observations* | 5,775 |
| *Number of firms* | 1,483 | *Number of firms* | 1,483 |
| $R^2$ | 0.076 | $R^2$ | 0.087 |

This table reports the estimated results from the regressions of the quality of green innovation on the total number of negative news reports of listed companies and the total number of negative news reports of listed companies on stock crash risk. t statistic based on the robust standard error is in parentheses. ***, ** and * indicate significance at the 1%, 5%, and 10% levels, respectively.

## 8| CONCLUSIONS AND DISCUSSIONS

Can green technological innovation affect the risk of a stock market crash? First, using a sample of Chinese listed companies from 2008 to 2018, we found that the quantitative indicator of green technology innovation is not related to the risk of a stock price collapse, while the quality indicator of green technology innovation is negatively related to the risk of a stock price collapse. Considering the noise in the Chinese stock market, the importance of the quality of green technology innovation in determining the risk of a stock market crash is worth noting. Next, we used the DID method to use the impact of external shock policies on green technological innovation to solve the endogeneity problem, and the main research results passed some robustness tests. Furthermore, we found that the negative correlation between the quality of green technology innovation and the risk of a stock price collapse is mainly reflected in nonstate-owned enterprises, while in state-owned enterprises, the quality of green technological innovation and the risk of a stock price collapse are positive and not significant. Finally, we found that the amount of negative news of listed companies played an intermediary role between the quality of green technology innovation and the risk of a stock price collapse. Specifically, the quality of green technology innovation is positively correlated with negative news reports of listed companies, and negative news reports of listed companies are, in turn, negatively correlated with the risk of a stock price collapse.

This research is an important paper to explore the impact of green technological innovation on stock price crashes. Unlike other factors that may affect the risk of stock crashes, activities that improve the quality of green technology innovation themselves have dual externalities (It has the characteristics of environment and innovative public goods). The empirical results show that the improvement of the quality of green technology innovation reflects the tendency of corporate shareholderism and restrains the risk of a stock price collapse. There are several reasons for this effect. First, the improvement of the quality of green technology innovation reflects whether corporate management is morally constrained against disclosing corporate information according to the real time window. Second, the improvement of the quality of green technology innovation has also improved the accuracy and timeliness of corporate information disclosure through contacts with relevant parties involved in the activities. In addition, we found that the media, as an important information channel, particularly the number of negative news reports involved in this study, plays an important role as a bridge between the quality of green technology innovation and the crash of stock prices; that is, management self-interest tendency and shareholderism is reflected in green technology innovation, affect the management's negative information hiding behavior and the company's information disclosure， which will ultimately affect the risk of a stock price crash by affecting the number of negative media reports.

Whether the behavior of green technological innovation can be fairly judged by the market and the virtuous circle mechanism that provides good feedback is worthy of attention. This can also effectively encourage enterprises to carry out green technological innovation, which is an innovation behavior that has certain risks due to the existence of dual externalities(Characteristics of environmental and innovative public goods). The empirical results show that the improvement of the quality of green technology innovation plays the role of a market stabilizer because it effectively suppresses the risk of a stock price collapse, and the positive feedback from the market also provides positive incentives for a company to continue to improve the quality of green technology innovation. For investors, the quality of green technological innovation, rather than the quantity, are important indicators that effectively assess the risk of a stock price crash and provide a new factor to be considered in the construction of the related asset management model. For government policymakers, when formulating financial policies, they need to pay attention to the disclosure of the quantity and quality indicators of green technology innovation, so as to guide and encourage enterprises to engage in green technology innovation and monitor the risk level of enterprises. For enterprise management, they should realize that adopting green innovation quality improvement strategy rather than quantity improvement strategy can bring good market feedback.

This paper selects Chinese listed companies as the research object. As China's capital market is still immature to a certain extent and the phenomenon of information asymmetry is more extensive, the corresponding regulatory measures still need to be improved (Huang, 2010;

Jiang and Kim, 2015). The impact of the quality improvement of green technology on stock price crashes is different between state-owned enterprises and nonstate-owned enterprises. For state-owned enterprises, improvements in the quality of green technology innovation are the embodiment of the decline of political resources. Related enterprises may adopt information-hiding behavior to reverse performance, which increases the risk of stock price crashes, in general, the special nature of state-owned enterprises distorts the original negative relationship between the quality of green technology innovation and the stock price collapse, in general, the relationship between the two is not significant, and there is a positive correlation to a certain extent. For nonstate-owned enterprises, the adoption of green technology innovation to improve quality reflects shareholder interests, which reduces the risk of stock price crashes. In addition, as China's environmental pollution is more serious than that of other countries, the impact of green technology innovation on stock crash risk may be more significant in China than in other countries. Due to the availability of data and workload, this paper does not conduct a transnational comparative analysis and looks forward to further exploring this dimension in future research.

## APPENDIX A. VARIABLE DEFINITIONS

This table contains the definitions of variables used in our analysis.

| | |
|---|---|
| $NCSKEW_{i,t+1}$ | Negative stock return skewness coefficient of company i in year t+1 |
| $DUVOL_{i,t+1}$ | Ratio of fluctuations in stock return of company i in year t+1 |
| $NCSKEW_{i,t}\left(DUVOL_{i,t}\right)$ | the negative return skewness coefficient and the ratio of fluctuations in the stock return of company i in year t |
| $NGInnovation1_{i,t}$ | Quantity index 1 of green patents of company i in year t |
| $NGInnovation2_{i,t}$ | Quantity index 2 of green patents of company i in year t |
| $QGInnovation_{i,t}$ | Quality index of green patents of company i in year t |
| $ONMEDIA_{i,t}$ | Number of original negative information reports of company i in year t |
| $ANMEDIA_{i,t}$ | Total number of negative reports of company i in year t |
| $Oturnover_{i,t}$ | The average monthly excess turnover rate is the average monthly turnover rate of stock i in year t - the average monthly turnover rate of stock i in year t-1 |
| $Sigma_{i,t}$ | Volatility in earnings. Standard deviation of stock i's mid-week yield in year t |
| $Ret_{i,t}$ | Average weekly yield. The average weekly return rate of stock in year t in year i |
| $Size_{i,t}$ | Natural logarithm of total assets of stock i in year t |
| $BM_{i,t}$ | Book to market ratio. Net assets of stock i in year t/(share price at the end of year t x number of circulating shares + net assets per share x number of non-circulating shares) |
| $Lev_{i,t}$ | Asset liability ratio = Total liabilities/total assets of stock i in year t |
| $Roa_{i,t}$ | Return on total assets = Net profit/total assets of stock i in year t |
| $ABACC_{i,t}$ | Index of the degree of information asymmetry of i stock in year t, equal to the absolute value of the residual of the correct Jones model |
| $Tophold_{i,t}$ | Shareholding ratio of the largest shareholder of company i in year t |
| $Power_{i,t}$ | The dummy variable of management power of i stock in year t, equal to 1 when the chairman and CEO are concurrently serving, otherwise it is 0. |
| $Maghold_{i,t}$ | The dummy variable of the management shareholding ratio of i stock in year t; when the management shareholding ratio is greater than the annual industry median, it is 1, otherwise it is 0. |
| $Balance_{i,t}$ | The dummy variable of equity checks and balance of company i in year t. Use the sum of the shareholding ratio of the second to fifth largest shareholders/the shareholding ratio of the first largest shareholder to measure equity checks and balance, when the equity checks and balance of company are greater than the annual industry median, it is 1, otherwise it is 0. |
| $Inshold_{i,t}$ | The dummy variable of institutional investor holdings of i stock in year t. When the proportion of institutional investors is greater than the annual industry median, it is 1, otherwise it is equal to 0. |
| $Firm_{i,t}$ | refers to Firm i in year t. |

## REFERENCES

Aupperle, K. E., Carroll, A. B., & Hatfield, J. D. (1985). An empirical examination of the relationship between corporate social responsibility and profitability. Academy of management

Journal, 28(2), 446-463.

An, H., & Zhang, T. (2013). Stock price synchronicity, crash risk, and institutional investors. Journal of Corporate Finance, 21, 1-15.

Andreou, P. C., Antoniou, C., Horton, J., & Louca, C. (2016). Corporate governance and firm-specific stock price crashes. European Financial Management, 22(5), 916-956.

Andreou, P. C., Louca, C., & Petrou, A. P. (2017). CEO age and stock price crash risk. Review of Finance, 21(3), 1287-1325.

Al Mamun, M., Balachandran, B., & Duong, H. N. (2020). Powerful CEOs and stock price crash risk. Journal of Corporate Finance, 62, 101582.

Barnea, A., & Rubin, A. (2010). Corporate social responsibility as a conflict between shareholders. Journal of business ethics, 97(1), 71-86.

Bu, M., & Wagner, M. (2016). Racing to the bottom and racing to the top: The crucial role of firm characteristics in foreign direct investment choices. Journal of International Business Studies, 47(9), 1032-1057.

Bhargava, R., Faircloth, S., & Zeng, H. (2017). Takeover protection and stock price crash risk: Evidence from state antitakeover laws. Journal of Business Research, 70, 177-184.

Comanor, W. S., & Scherer, F. M. (1969). Patent statistics as a measure of technical change. Journal of political economy, 77(3), 392-398.

Christmann, P. (2000). Effects of “best practices” of environmental management on cost advantage: The role of complementary assets. Academy of Management journal, 43(4), 663-680.

Chen, J., Hong, H., & Stein, J. C. (2001). Forecasting crashes: Trading volume, past returns, and conditional skewness in stock prices. Journal of financial Economics, 61(3), 345-381.

Chen, Y. S., Lai, S. B., & Wen, C. T. (2006). The influence of green innovation performance on corporate advantage in Taiwan. Journal of business ethics, 67(4), 331-339.

Choi, J., & Wang, H. (2009). Stakeholder relations and the persistence of corporate financial performance. Strategic management journal, 30(8), 895-907.

Callen, J. L., & Fang, X. (2013). Institutional investor stability and crash risk: Monitoring versus short-termism?. Journal of Banking & Finance, 37(8), 3047-3063.

Callen, J. L., & Fang, X. (2015). Short interest and stock price crash risk. Journal of Banking & Finance, 60, 181-194.

Chen, X., Huang, Q., & Zhang, F. (2015). CEO Duality and Stock Price Crash Risk: Evidence from China. Available at SSRN 2688779.

Chuang, S. P., & Huang, S. J. (2015). Effects of business greening and green IT capital on business competitiveness. Journal of Business Ethics, 128(1), 221-231.

Claudy, M. C., Peterson, M., & Pagell, M. (2016). The roles of sustainability orientation and market knowledge competence in new product development success. Journal of Product Innovation Management, 33, 72-85.

Cainelli, G., D’Amato, A., & Mazzanti, M. (2020). Resource efficient eco-innovations for a circular economy: Evidence from EU firms. Research Policy, 49(1), 103827.

Colombelli, A., Ghisetti, C., & Quatraro, F.(2020). Green technologies and firms’ market value: a micro-econometric analysis of European firms. Industrial and Corporate Change, 29(3), 855-875.

Dechow, P. M., Sloan, R. G., & Sweeney, A. P. (1995). Detecting earnings management. Accounting review, 193-225.

Dyck, A., & Zingales, L. (2004). Private benefits of control: An international comparison. The journal of finance, 59(2), 537-600.

Du, X. (2015). How the market values greenwashing? Evidence from China. Journal of Business Ethics, 128(3), 547-574.

DeFond, M. L., Hung, M., Li, S., & Li, Y. (2015). Does mandatory IFRS adoption affect crash risk?. The Accounting Review, 90(1), 265-299.

Friedman, M. (1970). VThe Social Responsibility of Business Is to Increase Its Profits. V The New York Times September, 13, 32-33.

Fombrun, C., & Shanley, M. (1990). What's in a name? Reputation building and corporate strategy. Academy of management Journal, 33(2), 233-258.

Freeman, R. E., & Liedtka, J. (1991). Corporate social responsibility: A critical approach. Business horizons, 34(4), 92-99.

Fombrun, C. J. (2005). A world of reputation research, analysis and thinking—building corporate reputation through CSR initiatives: evolving standards. Corporate reputation review, 8(1), 7-12.

Freeman, R. E., Harrison, J. S., & Wicks, A. C. (2007). Managing for stakeholders: Survival, reputation, and success. Yale University Press.

Fang, L., & Peress, J. (2009). Media coverage and the cross-section of stock returns. The Journal of Finance, 64(5), 2023-2052.

Francis, B., Hasan, I., & Li, L. (2016). Abnormal real operations, real earnings management, and subsequent crashes in stock prices. Review of Quantitative Finance and Accounting, 46(2), 217-260.

Gelb, D. S., & Strawser, J. A. (2001). Corporate social responsibility and financial disclosures: An alternative explanation for increased disclosure. Journal of Business Ethics, 33(1), 1-13.

Griliches, Z. (2007). 13. Patent Statistics as Economic Indicators: A Survey (pp. 287-344). University of Chicago Press.

Harvey, C. R., & Siddique, A. (2000). Conditional skewness in asset pricing tests. The Journal of finance, 55(3), 1263-1295.

Hillman, A. J., & Keim, G. D. (2001). Shareholder value, stakeholder management, and social issues: What's the bottom line?. Strategic management journal, 22(2), 125-139.

Hagedoorn, J., & Cloodt, M. (2003). Measuring innovative performance: is there an advantage in using multiple indicators?. Research policy, 32(8), 1365-1379.

Haigh, M., & Hazelton, J. (2004). Financial markets: a tool for social responsibility?. Journal of Business Ethics, 52(1), 59-71.

Hemingway, C. A., & Maclagan, P. W. (2004). Managers' personal values as drivers of corporate social responsibility. Journal of business ethics, 50(1), 33-44.

Hall, B. H., Jaffe, A., & Trajtenberg, M. (2005). Market value and patent citations. RAND Journal of economics, 16-38.

Hull, C. E., & Rothenberg, S. (2008). Firm performance: The interactions of corporate social performance with innovation and industry differentiation. Strategic management journal, 29(7), 781-789.

Hutton, A. P., Marcus, A. J., & Tehranian, H. (2009). Opaque financial reports, R2, and crash risk. Journal of financial Economics, 94(1), 67-86.

Huang, H. (2010). Institutional structure of financial regulation in China: lessons from the global financial crisis. Journal of Corporate Law Studies, 10(1), 219-254.

Ioannou, I., & Serafeim, G. (2015). The impact of corporate social responsibility on investment recommendations: Analysts' perceptions and shifting institutional logics. Strategic Management Journal, 36(7), 1053-1081.

Jin, L., & Myers, S. C. (2006). R2 around the world: New theory and new tests. Journal of financial Economics, 79(2), 257-292.

Jiang, F., & Kim, K. A. (2015). Corporate governance in China: A modern perspective.

Juntunen, J. K., Halme, M., Korsunova, A., & Rajala, R. (2019). Strategies for integrating stakeholders into sustainability innovation: a configurational perspective. Journal of Product Innovation Management, 36(3), 331-355.

Kapstein, E. B. 2001. The corporate ethics crusade. Foreign affairs, 105-119.

Kothari, S. P., Shu, S., & Wysocki, P. D. (2009). Do managers withhold bad news?. Journal of Accounting research, 47(1), 241-276.

Kim, J. B., Li, Y., & Zhang, L. (2011a). Corporate tax avoidance and stock price crash risk: Firm-level analysis. Journal of Financial Economics, 100(3), 639-662.

Kim, J. B., Li, Y., & Zhang, L. (2011b). CFOs versus CEOs: Equity incentives and crashes. Journal of financial economics, 101(3), 713-730.

Kim, Y., Park, M. S., & Wier, B. (2012). Is earnings quality associated with corporate social responsibility?. The accounting review, 87(3), 761-796.

Kim, J. B., & Zhang, L. (2014). Financial reporting opacity and expected crash risk: Evidence from implied volatility smirks. Contemporary Accounting Research, 31(3), 851-875.

Kim, Y., Li, H., & Li, S. (2014). Corporate social responsibility and stock price crash risk. Journal of Banking & Finance, 43, 1-13.

Kim, J. B., & Zhang, L. (2016). Accounting conservatism and stock price crash risk: Firm-level evidence. Contemporary accounting research, 33(1), 412-441.

Kim, J. B., Wang, Z., & Zhang, L. (2016). CEO overconfidence and stock price crash risk. Contemporary Accounting Research, 33(4), 1720-1749.

Lewis, A., & Juravle, C. (2010). Morals, markets and sustainable investments: A qualitative study of 'champions'. Journal of Business Ethics, 93(3), 483-494.

Luo, J. H., Gong, M., Lin, Y., & Fang, Q. (2016). Political connections and stock price crash risk: Evidence from China. Economics Letters, 147, 90-92.

Li, X., Wang, S. S., & Wang, X. (2017). Trust and stock price crash risk: Evidence from China. Journal of Banking & Finance, 76, 74-91.

Li, J., Wang, L., Zhou, Z. Q., & Zhang, Y. (2021). Monitoring or tunneling? Information interaction among large shareholders and the crash risk of the stock price. Pacific-Basin Finance Journal, 65, 101469.

Leonidou, L. C., Christodoulides, P., Kyrgidou, L. P., & Palihawadana, D. (2017). Internal drivers and performance consequences of small firm green business strategy: The moderating role of external forces. Journal of business ethics, 140(3), 585-606.

Leyva-de la Hiz, D. I., Ferron-Vilchez, V., & Aragon-Correa, J. A. (2019). Do firms' slack resources influence the relationship between focused environmental innovations and financial performance? More is not always better. Journal of Business Ethics, 159(4), 1215-1227.

McWilliams, A., & Siegel, D. (2000). Corporate social responsibility and financial performance: correlation or misspecification?. Strategic management journal, 21(5), 603-609.

Marin, G., & Lotti, F. (2017). Productivity effects of eco-innovations using data on eco-patents. Industrial and corporate change, 26(1), 125-148.

Nelling, E., & Webb, E.(2009). Corporate social responsibility and financial performance: the "virtuous circle" revisited. Review of Quantitative Finance and Accounting, 32(2), 197-209.

Ni, X., & Zhu, W. (2016). Short-sales and stock price crash risk: Evidence from an emerging market. Economics Letters, 144, 22-24.

Pastena, V., & Ronen, J. (1979). Some hypotheses on the pattern of management's informal disclosures. Journal of Accounting Research, 550-564.

Pujari, D., Wright, G., & Peattie, K. (2003). Green and competitive: Influences on environmental new product development performance. Journal of business Research, 56(8), 657-671.

Pujari, D. (2006). Eco-innovation and new product development: understanding the influences on market performance. Technovation, 26(1), 76-85.

Petrovits, C. M. (2006). Corporate-sponsored foundations and earnings management. Journal of Accounting and Economics, 41(3), 335-362.

Peress, J. (2014). The media and the diffusion of information in financial markets: Evidence from newspaper strikes. The Journal of Finance, 69(5), 2007-2043.

Stanwick, P. A., & Stanwick, S. D. (1998). The relationship between corporate social performance, and organizational size, financial performance, and environmental performance: An empirical examination. Journal of business ethics, 17(2), 195-204.

Simpson, W. G., & Kohers, T. (2002). The link between corporate social and financial performance: Evidence from the banking industry. Journal of business ethics, 35(2), 97-109.

Scandelius, C., & Cohen, G. (2016). Achieving collaboration with diverse stakeholders—The role of strategic ambiguity in CSR communication. Journal of Business Research, 69(9), 3487-3499.

Stucki, T. 2019. Which firms benefit from investments in green energy technologies?–The effect of energy costs. Research Policy, 48(3), 546-555.

Shahab, Y., Ntim, C. G., Ullah, F., Yugang, C., & Ye, Z. (2020). CEO power and stock price crash risk in China: Do female directors' critical mass and ownership structure matter?. International Review of Financial Analysis, 68, 101457.

Tien, S. W., Chung, Y. C., & Tsai, C. H. (2005). An empirical study on the correlation between environmental design implementation and business competitive advantages in Taiwan's industries. Technovation, 25(7), 783-794.

Tong, Z., Chen, Y., Malkawi, A., Liu, Z., & Freeman, R. B. (2016). Energy saving potential of natural ventilation in China: The impact of ambient air pollution. Applied energy, 179, 660-668.

Wang, Q., Wong, T. J., & Xia, L. (2008). State ownership, the institutional environment, and auditor choice: Evidence from China. Journal of accounting and economics, 46(1), 112-134.

Weche, J. P. (2019). Does green corporate investment crowd out other business investment?. Industrial and Corporate Change, 28(5), 1279-1295.

Xu, N., Jiang, X., Chan, K. C., & Yi, Z. (2013). Analyst coverage, optimism, and stock price crash risk: Evidence from China. Pacific-Basin Finance Journal, 25, 217-239.

Xu, N., Li, X., Yuan, Q., & Chan, K. C. (2014). Excess perks and stock price crash risk: Evidence from China. Journal of Corporate Finance, 25, 419-434.

Xie, X., Huo, J., & Zou, H. (2019). Green process innovation, green product innovation, and corporate financial performance: A content analysis method. Journal of Business Research, 101, 697-706.

Ye, K., & Zhang, R. (2011). Do lenders value corporate social responsibility? Evidence from China. Journal of Business Ethics, 104(2), 197-206.

Yuan, R., Sun, J., & Cao, F. (2016). Directors' and officers' liability insurance and stock price crash risk. Journal of Corporate Finance, 37, 173-192.

Zaman, R., Atawnah, N., Haseeb, M., Nadeem, M., & Irfan, S. (2021). Does corporate eco-innovation affect stock price crash risk?. The British Accounting Review, 101031.

Zhang, R., Zhu, J., Yue, H., & Zhu, C. (2010). Corporate philanthropic giving, advertising intensity, and industry competition level. Journal of business Ethics, 94(1), 39-52.

Zhang, M., Ma, L., Su, J., & Zhang, W. (2014). Do suppliers applaud corporate social performance?. Journal of Business Ethics, 121(4), 543-557.

Zhang, M., Xie, L., & Xu, H. (2016). Corporate philanthropy and stock price crash risk: Evidence from China. Journal of Business Ethics, 139(3), 595-617.

Zhang, M., Liu, Y., Xie, L., & Ye, T. (2017). Does the cutoff of “red capital” raise a red flag? Political connections and stock price crash risk. The North American Journal of Economics and Finance, 39, 89-109.

Zhang, L., Cao, C., Tang, F., He, J., & Li, D. (2019). Does China’s emissions trading system foster corporate green innovation? Evidence from regulating listed companies. Technology Analysis & Strategic Management, 31(2), 199-212.